\documentclass[aps,prx,twocolumn,longbibliography,superscriptaddress]{revtex4-2}
\usepackage{graphicx}
\usepackage{dcolumn}
\usepackage{bm}
\usepackage[citecolor=blue, colorlinks=true, urlcolor=blue, linkcolor=blue]{hyperref}
\usepackage{xcolor}
\usepackage{gensymb}
\usepackage{listings} 

\usepackage{physics}
\usepackage{todonotes}
\usepackage{subfiles}
\usepackage{amsmath}
\usepackage{amssymb}
\usepackage{mathtools}
\usepackage{float}
\usepackage{braket}
\usepackage{svg}
\usepackage{wasysym}   

\definecolor{codegreen}{rgb}{0,0.6,0}
\definecolor{codegray}{rgb}{0.5,0.5,0.5}
\definecolor{codepurple}{rgb}{0.58,0,0.82}
\definecolor{backcolour}{rgb}{0.95,0.95,0.92}

\lstdefinestyle{mystyle}{
    backgroundcolor=\color{backcolour},   
    commentstyle=\color{codegreen},
    keywordstyle=\color{magenta},
    numberstyle=\tiny\color{codegray},
    stringstyle=\color{codepurple},
    basicstyle=\ttfamily\footnotesize,
    breakatwhitespace=false,         
    breaklines=true,                 
    captionpos=b,                    
    keepspaces=true,                 
    numbers=left,                    
    numbersep=5pt,                  
    showspaces=false,                
    showstringspaces=false,
    showtabs=false,                  
    tabsize=2
}

\begin{document}

\newcommand{\TUM}{\affiliation{Technical University of Munich, TUM School of Natural Sciences, Physics Department, 85748 Garching, Germany}}
\newcommand{\MCQST}{\affiliation{Munich Center for Quantum Science and Technology (MCQST), Schellingstr. 4, 80799 M{\"u}nchen, Germany}}

\author{Gloria Isbrandt}\TUM \MCQST
\author{ Frank Pollmann}\TUM \MCQST
\author{Michael Knap} \TUM \MCQST

\date{\today}

\title{Anisotropic Spin Ice on a Breathing Pyrochlore Lattice}

\begin{abstract}
Spin ice systems represent a prime example of constrained spin systems and exhibit rich low-energy physics. In this study, we explore how introducing a tunable anisotropic spin coupling to the conventional Ising spin ice Hamiltonian on the breathing pyrochlore lattice affects the ground state properties of the system. Significant changes are observed in the ground state structure, reflected in the spin structure factor and in a reduction of residual entropy at low temperatures.  
We theoretically uncover a rich phase diagram by varying the anisotropy and demonstrate how this modification reduces the ground state degeneracy across different phases. Numerical simulations reveal that, at sufficiently low temperatures, the system either undergoes a crossover into a constrained spin ice manifold, characterized by an entropy density that drops below the Pauling entropy of conventional spin ice, or a phase transition into a symmetry-broken state, depending on the perturbations. Additionally, we compute the spin structure factors for the anisotropic model and compare these results to analytical predictions from a self-consistent Gaussian approximation, finding good agreement.  
This work develops the understanding of spin ice in anisotropic limits, which may be experimentally realized by strain, providing, among others, key signatures in entropy and specific heat.

\end{abstract}

\maketitle

\section{Introduction}  
Spin ice systems have been a prominent focus of condensed matter research since their discovery~\cite{Harris1997Sep, Bramwell1998Apr}, due to their exotic ground states~\cite{Ramirez1999May} and the emergence of quasiparticle excitations resembling magnetic monopoles~\cite{Castelnovo2008Jan, Castelnovo2012Mar}. These systems embody the interplay of geometric frustration and local constraints, resulting in a macroscopically degenerate ground state governed by the so-called ``ice rule." This rule dictates the arrangement of spins on a tetrahedron in the pyrochlore lattice, requiring that two spins point into the tetrahedron, aligned along the local cubic [111] directions, and two spins point out~\cite{Bramwell1998Apr, Moessner1998Mar}.  

While the simple nearest-neighbor spin-ice model is by now well understood~\cite{Bramwell1998Apr, Moessner1998Mar, Moessner2021Oct, Gingras2021Oct}, several questions arise when comparing it to experiments. 
Experimental studies of Dy$_2$Ti$_2$O$_7$ have revealed, for instance, that thermally equilibrated samples, cooled over long timescales, exhibit an entropy drop below the celebrated Pauling entropy, a phenomenon that remains under active investigation~\cite{Pomaranski2013Jun, Bermudez2025Apr}. One possible explanation is the formation of a quantum spin liquid, where the true ground state would be a coherent superposition of the classical configurations, thereby lifting the extensive degeneracy~\cite{moessner2003, Hermele2004Feb, Shannon2012Feb}. Another explanation for the release of residual entropy is based on further neighbor or long-range interactions present in spin ice materials. These naturally change the degeneracy of the ground state and lead to an ordered state at low energies~\cite{Melko2001Jul, Melko2004Oct, Yavorskii2008Jul}.
An alternative approach to release the residual entropy has been shown to be through strain engineering. Through theoretical~\cite{ Jaubert2010Aug, Jaubert2017May} and experimental studies in spin ice thin films~\cite{Bovo2014Mar, Bovo2019Mar}, it has been shown that spin ice materials, when put under strain along certain lattice directions, the system's physics will change at low temperatures, degeneracies are lifted, and the third law of thermodynamics is restored. 
Beyond the spin ice models on the pyrochlore lattice, the breathing pyrochlore lattice—a structural variant of the conventional pyrochlore lattice—has emerged as a promising platform for studying novel phases of matter. In this lattice, alternating tetrahedra of different sizes and bond strengths break the inherent symmetry of the system, strongly influencing its magnetic properties~\cite{Rau2016Jun, Bag2023Apr, Ghosh2019}. This structural feature has been linked to novel phenomena, including fracton physics and the potential realization of distinct quantum spin liquid phases~\cite{Yan2020Mar, Han2022Jun, Sanders2024Mar}. 
This leads to the question of how the anisotropic interactions of a strained system would change the behavior if we allowed for a distinction of the tetrahedral sublattices, as is the case with a breathing pyrochlore lattice. 
While uniform perturbations, such as external magnetic fields, further neighbor interactions, or global strain, have been extensively explored~\cite{Lu2024Jun, Jaubert2010Aug, Jaubert2017May, Jaubert2008Feb, Hallen2024Jan}, the effects of spatially dependent modifications remain open. 

In this work, we introduce an anisotropic perturbation: a bond-dependent tuning parameter that effectively acts as a sublattice-dependent strain along the $[001]$ direction. This parameter enables independent tuning of the interaction strengths on the two tetrahedral sublattices of the breathing pyrochlore lattice, breaking the lattice symmetry in a controlled and tunable manner.  
Strain engineering in spin-ice materials, such as Dy$_2$Ti$_2$O$_7$ and Ho$_2$Ti$_2$O$_7$, has been experimentally realized in various studies~\cite{Bovo2019Mar, Bovo2014Mar, Barry2019Aug, Jaubert2010Aug}. However, these studies have primarily focused on uniform strain. Here, we go beyond this by providing a comprehensive theoretical analysis of sublattice-dependent strain on a toy model based on the breathing pyrochlore lattice. Using Monte Carlo simulations, we explore the phases that arise in the ground-state limit for different combinations of strain parameters and analyze the system's equilibrium properties. 
Interestingly, we find that in the ground state limit, for different strains, the system effectively decouples into lower-dimensional subsystems, changing the ground state degeneracies. This dimensional reduction can be understood in the framework of \textit{intermediate} symmetries, where symmetry operations act on lower-dimensional subregions, such as lines or planes, within a higher-dimensional system~\cite {Nussinov2015Jan}. While global symmetries act on the whole system and local (gauge) symmetries on single points in space, intermediate symmetries apply uniformly to $d$-dimensional subspaces in a $D$-dimensional system. These symmetries can lead to dimensional reduction through effective energetic decoupling of subsystems~\cite{Nussinov2009Oct, Nussinov2012Oct, Nussinov2015Jan}. 

From a different perspective, this decoupling mechanism is inherently tied to frustration. In the unperturbed spin ice model, all bonds are equally frustrated; introducing anisotropy lifts frustration selectively, favoring certain local configurations and decoupling the system into independent planes or lines. A classical example of this mechanism is found in the triangular lattice Ising antiferromagnet, where increasing anisotropy along one lattice direction reduces frustration and leads to decoupled one-dimensional Ising chains at $T = 0$~\cite{Stephenson1970Feb}. By contrast, in unfrustrated three-dimensional systems, anisotropy typically leads only to a crossover in thermodynamic behavior, not full decoupling, since finite coupling persists between subsystems~\cite{Sengupta2003Sep, KyuWonLee2002, Borel2025Apr}. Thus, frustration plays a crucial role in enabling genuine dimensional reduction in our model.
Our work provides a comprehensive understanding of how lattice asymmetry can be leveraged to engineer the ground state and thermodynamic properties of constrained spin systems and generate dimensional reduction in a frustrated system. 

This paper is structured as follows: First, we introduce the model in Sec.~\ref{sec:Toy_Model} and describe the types of strain investigated. In Sec.~\ref{sec:GSD}, we derive the ground state degeneracies for the different strain cases. We then analyze the finite-temperature thermodynamics of the model in Sec.~\ref{sec:Thermodynamics}, using numerical Monte Carlo simulations to study specific heat and entropy. In Sec.~\ref{sec:SSF}, we compare numerical calculations of the spin structure factor with analytical results obtained in the large-$N$ limit. We provide an outlook in Sec.~\ref{sec:Summary} and relegate technical details to the appendix.

\section{Model}
\label{sec:Toy_Model}

\begin{figure}[h!]
    \centering
    \includegraphics[width = 1 \columnwidth]{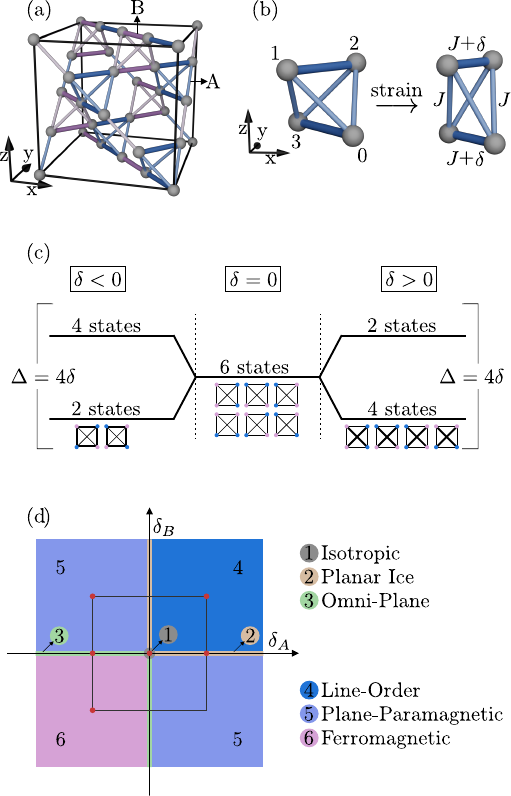}
    \caption{\textbf{Anisotropic breathing pyrochlore.} (a) Unit cell of the breathing pyrochlore lattice with the different tetrahedral sublattices shown in two different colors. The strained bonds (thick tubes) lie in the $x,y$-plane of the unit cell. (b) Single tetrahedron with numbering convention of the lattice sites. The strained tetrahedron is shown with the corresponding interaction coefficients. (c) The strain $ \delta$ lifts the degeneracy of the six ground states of a single tetrahedron. For $\delta<0$, a single tetrahedron has two ground states, while for $\delta>0$, a single tetrahedron has four ground states. (d) Phase diagram with six different phases depending on the sign of $\delta_{A/B}$. The mid-point, for $\delta_A = \delta_B = 0$, is the conventional isotropic spin ice model.}
    \label{fig:Lattice}
\end{figure}

We consider a pyrochlore lattice, shown in Fig. \ref{fig:Lattice}(a), with interacting spins residing at the lattice sites shown in grey. We assume the system is in the spin ice limit, induced by a large crystal field anisotropy that constrains the spins to point along the local cubic $[111]$ direction~\cite{Bramwell1998Apr, Moessner1998Mar}.
In rare-earth pyrochlore spin ice materials such as Dy$_2$Ti$_2$O$_7$ and Ho$_2$Ti$_2$O$_7$, the magnetic interactions consist of two main contributions: a short-range superexchange interaction $J_{\text{nn}}$, typically antiferromagnetic, and a long-range dipolar interaction $D_{\text{nn}}$, which is ferromagnetic. Together, these define an effective nearest-neighbor coupling $J_{\text{eff}} = J_{\text{nn}} + D_{\text{nn}}$, which is positive for the materials mentioned and stabilizes the spin ice regime by enforcing the "2-in, 2-out" ice rule at intermediate temperatures~\cite{Bramwell2001Jul, Melko2001Jul}. In this work, we adopt a minimal classical model that retains only the nearest-neighbor exchange interaction, effectively truncating dipolar interactions beyond first neighbors. This simplification maps the system onto a classical Ising problem with antiferromagnetic nearest-neighbor interactions, which remains a good approximation at temperatures above the scale set by the bandwidth introduced by dipolar terms~\cite{Isakov2004Oct, DenHertog2000Apr}. 
We additionally assume a distinction between the tetrahedral sublattices, in Fig.~\ref{fig:Lattice}(a), shown in different colors, which we will refer to as $A$ and $B$ sublattice. This distinction is typically applicable to the breathing pyrochlore lattice~\cite{Rau2016Jun,Han2022Jun, Sadoune2024Feb}, where, for example, due to different atoms, the size of the tetrahedra of the two sublattices is different. As a consequence, our interaction constants can be individually tuned on the sublattices.
In our model, however, we want to make an additional change to the commonly known spin-ice problem: we will add an additional perturbation that increases or decreases the interaction between two specific spin pairs, $\{ S_0^z, S_3^z\}$ and  $\{ S_1^z, S_2^z\}$,  by $\delta_{A/B}$ on both sublattices independently and making it thereby anisotropic. This is shown in Fig.~\ref{fig:Lattice}(a) and (b). 

A possible realization may be an anisotropic pressure-induced strain on the lattice, where the pressure is applied along the $[001]$ direction for $\delta<0$ or onto the plane perpendicular to it for $\delta >0$, thereby reducing or increasing the distance between the lattice sites and making the unit cell tetragonal or orthorhombic. An example of a strained single tetrahedron in $[110]$ direction is shown in Fig. \ref{fig:Lattice}(b). 
Pressure-induced strain and change of the behavior of spin ice materials have been shown previously in single crystals and epitaxially strained thin films~\cite{Barry2019Aug, Mito2007Mar, Edberg2020Nov, Bovo2014Mar, Bovo2019Mar}. For a detailed description of possible experimental realizations, we refer to Appendix \ref{sec:App_Experimental_considerations}.

Our interaction Hamiltonian on the two different sublattices now looks as follows:
\begin{equation}
\label{Htot}
\begin{split}
    H &=  \sum_{\langle i,j \rangle \in A} J_A^{ij} S_i^zS_j^z + \sum_{\langle i,j \rangle \in B} J_B^{ij} S_i^zS_j^z, \\
\end{split}
\end{equation}
with the interaction coefficicent $J_{\alpha}^{ij} = J_{\alpha}+ \delta_{\alpha}$ for bonds in the $x,y$-plane, or $ij \in \{03, 12\}$, and otherwise just $J_{\alpha}^{ij} = J_{\alpha}$ for all other bonds, as indicated in Fig.~\ref{fig:Lattice}(b). 
The chosen numbering of the spins on the tetrahedra is such that the spin pairs with different interaction, $\{ S_0^z, S_3^z\}$ and  $\{ S_1^z, S_2^z\}$, are lying in directions $[110]$ and $[-110]$, or, the spins in the $x,y$-plane of the unit cell. 

We now rewrite the Hamiltonian in terms of tetrahedral monopole charges $\mathsf{Q}_{A/B}$ and  partial tetrahedral charge $\mathsf{q}^{12}_{A/B}$ and  $\mathsf{q}^{03}_{A/B}$, with their corresponding definition:
\begin{equation}
    \begin{split}
        \mathsf{q}_{A/B}^{03} &= (S_0^z + S_3^z)\\
        \mathsf{q}_{A/B}^{12} &= (S_1^z + S_2^z)\\
        \mathsf{Q}_{A/B} &= (S_0^z + S_1^z + S_2^z +S_3^z) = \mathsf{q}_{A/B}^{03} + \mathsf{q}_{A/B}^{12},
    \end{split}
\end{equation}
with which the Hamiltonian now reads

\begin{equation}
\label{eq:Ham_Q}
    \begin{split}
        H = & \frac{J_A}{2} \sum_A \mathsf{Q}_A^2 + \frac{\delta_A}{2} \sum_A \left((\mathsf{q}_A^{03})^2 +  (\mathsf{q}_A^{12})^2 \right)  + \\
        & \frac{J_B}{2} \sum_B \mathsf{Q}_B^2 + \frac{\delta_B}{2} \sum_B \left((\mathsf{q}_B^{03})^2 +  (\mathsf{q}_B^{12})^2 \right) \\
        &-2( J_A+\delta_A) N_{\text{A-tet}} - 2(J_B +\delta_B) N_{\text{B-tet}} .
    \end{split}
\end{equation}

Since we are working with Ising spins, and for computational purposes, we used $|S| = 1$. Depending on the length of the spin, the last term of the equation will always read $- \frac{(J_\alpha +\delta_{\alpha})}{2}N_{\alpha-\text{tet}} 4|S|^2$, where $N_{\alpha-\text{tet}} = \frac{1}{4} N_{\text{spin}}$ is the number of tetrahedra in the system. From here on, we will neglect this constant energy shift and set the last term in eq.~\eqref{eq:Ham_Q} to zero.
We can distinguish six different cases by tuning $\delta_{A/B}$ to be either positive, negative, or zero.
The convenient form of the Hamiltonian in Eq.~\eqref{eq:Ham_Q} makes the ground-state structure apparent: since we assume $J_{A/B}>0$, a ground state must fulfill $ \mathsf{Q}_{A/B} =0$. However, depending on the sign of $\delta_{A/B}$ the energy is minimized by either minimizing $\mathsf{q}_{A/B}^2$ -- corresponding to $\mathsf{q}_{A/B} = 0$ -- or maximizing the latter by $\mathsf{q}_{A/B} = \pm 2$. 

For a single tetrahedron, with $\delta_{\alpha} = \delta = 0$, there are $6$ possible ground states on a single tetrahedron; all states that follow the rule ``\textit{two in - two out}" and have $ \mathsf{Q}_{A/B} =0$.  Introducing the $\delta$-term splits these $6$ states into a group of $4$  ($\mathsf{q}_{A/B} = 0$ ) and a group of $2$ ($\mathsf{q}_{A/B} = \pm 2$) states. Which of these is the ground state and which is the higher-lying state depends on the sign of $\delta$; see Fig. \ref{fig:Lattice}(c).

The sign of $\delta$ will also influence the total number of ground states the whole system will have, as well as the possible mobile excitations on top of this ground state. We will introduce the nomenclature listed in Tab.~\ref{tab:nomenclature} for the six cases inspired by the type of order in the ground state. From now on, we will assume $\delta_i < J_i$. The phase diagram with the six different phases arising from tuning $\delta_A$ and $\delta_B$ is shown in Fig. \ref{fig:Lattice}(d). 

\begin{table*}[t]
    \centering
    \begin{tabular}{c @{\hskip 0.5in} c@{\hskip 0.5in} c@{\hskip 0.5in} c }
    \hline \hline
    & name & parameter & GSD\\ \hline
        1. & \textbf{Isotropic} & $\delta_A = \delta_B = 0$ & $\sim (3/2)^{8 L_x L_y L_z}$ \\
        2. & \textbf{Planar Ice} & $\delta_A > 0$ and $ \delta_B = 0$& $(4/3)^{6 L_x L_y L_z}$ \\
        3. & \textbf{Omni-Plane} & $\delta_A < 0$ and $ \delta_B = 0$ & $ 2^{2L_{x}}+2^{2L_{y}}+2^{2L_{z}}-6$ \\
        4. & \textbf{Line-Order} & $\delta_A, \delta_B > 0$&  $2^{4 \cdot \text{GCF}(L_x, L_y) \cdot L_z }$ \\
        5. & \textbf{Plane-Paramagnetic} & $\delta_A>0$ and $ \delta_B < 0$& $ 2^{2 \cdot L_z}$ \\
        6. & \textbf{Ferromagnetic} & $\delta_A, \delta_B < 0$& $2$ \\
    \end{tabular}
    \caption{Six limits of the strain parameters with the corresponding ground state degeneracy (GSD). The model is symmetric under the exchange of indices $A$ and $B$. Here, $L_{\alpha}$ is the number of unit cells in one spatial dimension $\alpha \in (x,y,z)$; $N= 16 \cdot L_x\cdot L_y \cdot L_z$ is the total number of spins in the lattice.}
    \label{tab:nomenclature}
\end{table*}

Our goal is to describe the thermodynamic properties of the breathing spin ice in the different strain limits: their theoretical ground state degeneracy and mobility of excitations on top of the ground state, as well as experimental signatures, such as the specific heat and entropy density at different temperatures. 

\section{Ground state degeneracy}
\label{sec:GSD}

The number of ground states on a single tetrahedron decreases for any $\delta \neq 0$. In isotropic spin ice, $\delta = 0$, we only distinguish between 6 ground states (``\textit{two in - two out}"), 8 single-monopole states (``\textit{three in - one out}" or vice versa), and two double-monopoles (``\textit{all in}" or ``\textit{all out}"). For a single tetrahedron, a total magnetization vector can be defined, $\mathbf{M}_i = \sum_{\alpha} S^z_{i, \alpha} \mathbf{e}_{\alpha}$, where $\mathbf{e}_{\alpha}$ points along the local $z$ axis of the spin ($\alpha$) position in the tetrahedron ($i$) (see App.~\ref{Sec:App_LocalSpinBasis}). All six ground states of a single tetrahedron have a non-zero magnetization vector; their magnetization vector will point parallel or antiparallel to one of the three spatial dimensions $(x,y,z)$. Introducing the $\delta$-term energetically splits the six ground states into four lower-lying and two higher-lying states, for $\delta > 0$, and two lower-lying and two higher-lying states for $\delta<0$, as indicated in Fig.~\ref{fig:Lattice}(c).

\begin{figure}
    \centering
    \includegraphics[width = 1 \columnwidth]{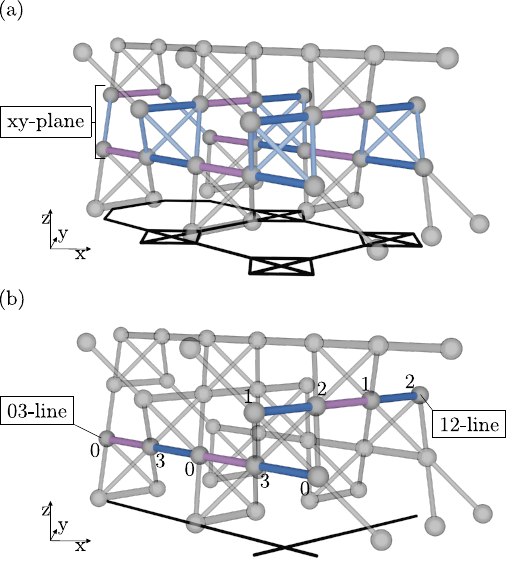}
    \caption{\textbf{Tetrahedra-planes and spin-lines.} (a) A plane of tetrahedra of the same kind is shown. Here, $A$ tetrahedra (blue) with equal $z$ coordinate, or an $x,y$-plane, are depicted as an example. The $A$ tetrahedra and the connecting links on the $B$ tetrahedra (lavender) are shown in color. The two-dimensional projection onto the plane is shown in black. Planes can be formed perpendicular to all spatial directions by either $A$ or $B$ tetrahedra. (b) Two spin lines are shown in color. The 03-line connects spins $S_0^z$ and $S_3^z$ in an $x,y$-plane and the 12-line connects spins $S_1^z$ and $S_2^z$. }
    \label{fig:Lines_Planes}
\end{figure}

For $\delta > 0$, the ground states have in common that the partial charge $\mathsf{q}^{03/12} = 0$, so the spin-pairs $S_0^z = -S_3^z$ and $S_1^z = -S_2^z$ and their magnetization vector points along the $\mathbf{M}_i \parallel \hat{x}$ or $\mathbf{M}_i \parallel \hat{y}$ direction, while for $\delta < 0$, the partial charges $\mathsf{q}^{03} = \mathsf{q}^{12} = \pm 2$ and the opposite, $S_0^z = S_3^z$ and $S_1^z = S_2^z$, holds, leading to $\mathbf{M}_i \parallel \hat{z}$. In any case, however, the total spin sum $\mathsf{Q}_{A/B} = \sum_i S_i^z = 0$.
From this, we can now build up the ground state for the whole system, matching tetrahedra from the $A$ and the $B$ sublattices. The number of ground states for all systems will decrease compared to spin ice. Importantly, the magnitude of $\delta_{A/B}$ is irrelevant when only considering the ground state degeneracy of the theoretical model. The magnitude will become relevant in experiments as well as when considering thermodynamic properties. 
Also, the energetic order of possible excitations changes depending on $\delta$, as listed in Tab.~\ref{tab:excitations_table}. 

In the following, we will analyze how introducing the $\delta$-terms alters the ground state degeneracy for the whole system. We will first determine how many configurations are ground states on a single tetrahedron of one species ($A/B$) in six different limits and then see how this relates to constructing a ground state on the whole lattice. Importantly, we will also discuss the operator structure necessary to construct all possible ground states. This will then relate to the name chosen for the different phases.
Lastly, we will also mention the mobility of excitations on top of the ground state. This will make apparent a connection to fracton physics, characterized by quasiparticle excitations being restricted in their mobility~\cite{Chamon2005Jan, Haah2011Apr, Yoshida2013Sep, Bravyi2013Nov, Vijay2015Dec, Yan2020Mar, Han2022Jun}.

\begin{table}[h]
    \centering
    \begin{tabular}{c c c}
    \hline \hline
    total charge & partial charge & energy\\ \hline
     $\mathsf{Q} = 0$&  $\mathsf{q}^{03} = \mathsf{q}^{12} = 0$  & $ \varepsilon = 0$  \\
    $\mathsf{Q} = 0$&  $\mathsf{q}^{03} = -\mathsf{q}^{12} = \pm 2$  & $\varepsilon = 4\delta$ \\
    $\mathsf{Q} = \pm2$&  $\mathsf{q}^{03} = \pm2 $ or $\mathsf{q}^{12} = \pm 2$  & $\varepsilon = 2J+2\delta$ \\
    $\mathsf{Q} = \pm4$&  $\mathsf{q}^{03} = \mathsf{q}^{12} = \pm 2$  & $\varepsilon = 8J + 4\delta $ \\ \hline
    \end{tabular}
    \caption{Energy of a single tetrahedron for the 4 different types of excitations. Which of these states are ground states depends on the sign of $\delta$.}
    \label{tab:excitations_table}
\end{table}

\subsection{Isotropic Spin Ice, $\delta_A = \delta_B = 0$}
\label{Ssec:Isotropic}

In the well-known case $\delta_A = \delta_B = 0$, the ground state of each tetrahedron is six-fold degenerate. It is characterized by a total spin-charge of $\mathsf{Q} = 0$ on each tetrahedron. This is the classical nearest-neighbor spin ice case.
The number of ground states for classical spin ice can be estimated analogously to the estimation done by Pauling for water ice~\cite{Pauling1935Dec}; the total number of possible configurations $2^N$ for all spins times the fraction of allowed configurations for each tetrahedron $(\frac{6}{16})^{N/2}$ gives an estimate for the number of ground states $\Omega = (3/2)^{N/2}$. Here, $N$ is the number of spins in the system. Equivalently, we will use $L_{\alpha}$ as the number of unit cells in spatial direction $\alpha$; each unit cell containing four tetrahedra of the same sublattice (see Fig.~\ref{fig:Lattice}(a)). Since on each tetrahedron, there are $4$ spins, and we can rewrite $\Omega = (3/2)^{8 L_x L_y L_z}$. Starting from any ground state configuration for spin ice, any other configuration can be reached by forming a closed loop of spins aligning head to tail (or ``\textit{in}" and  ``\textit{out}" alternatingly) and flipping them all. Since no anisotropies are at play here and the system possesses the full lattice symmetry, we will refer to this phase as \textit{isotropic} spin ice.
Flipping a single spin creates two monopole charges $\mathsf{Q} = \pm1$. In our nearest neighbor model, single monopoles can move to neighboring tetrahedra by flipping a single spin (of the three equally aligned) at zero energy cost \cite{Castelnovo2008Jan}. Here, the path of the monopoles, ``\textit{three in - one out}" or opposite, is effectively only constrained by the position of the ``\textit{one}" not equally aligned spin. The monopole diffusion can be considered three-dimensional since there is no actual favoring of certain alignments of spins. This picture only changes when including additional energetic considerations that lead to percolation into a fractal structure where monopoles can move~\cite{Stern2021Sep}.

While the here-assumed nearest neighbor spin ice model can explain many features and experimental signatures of spin ice, the long-range, dipolar interactions between spins are truncated beyond the nearest neighbors. Theoretical investigations have shown that when considering long-range interaction, further ordering at very low temperatures occurs. However, direct experimental observations have been proven challenging due to the low temperatures required for these features to be visible~\cite{Siddharthan1999Aug, Melko2001Jul, Gingras2011Feb, Melko2004Oct}. We will now show that in our anisotropic model, the ground state degeneracy always drops, even without considering the long-range dipolar interactions.

\subsection{Planar Ice, $\delta_A >0, \delta_B = 0$}
\label{Ssec:PlanarIce}
For $\delta_A > 0$ and $\delta_B = 0$, we distinguish between the $A$ and $B$ sublattice. On the $A$ sublattice, the total charge must be $\mathsf{Q}_A = 0$ and the partial charge $\mathsf{q}_A^{03/12} = 0$, while on the $B$ sublattice only the total charge must be $\mathsf{Q}_B = 0$; the value of the partial charge is not fixed.
So, on the $A$-tetrahedra, 4 ground states are allowed, while on the $B$-tetrahedra, all 6 spin-ice ground states are allowed. At close inspection, one notices that within single $x,y$-planes of $B$ tetrahedra, as indicated in Fig.~\ref{fig:Lines_Planes}, the system actually maps exactly to the square ice model (six vertex model or, equivalently, fully packed loop model). For this reason, we will refer to this phase as \textit{planar ice}.
The ground state degeneracy in the six vertex model on the square lattice can be computed exactly~\cite{Lieb1967Oct} and is given by $\Omega = \left(\frac{4}{3} \right)^{3/2 \cdot N_{\text{vert}}}$, where $N_{\text{vert}}$ are the number of vertices. In our system, the number of vertices is the number of $B$-tetrahedra in the strained plane ($2 \cdot L_x \cdot L_y$). Since the ground states can be chosen independently on each plane, and there are $2L_z$ planes within the system, the ground state degeneracy scales as $\left(\frac{4}{3} \right)^{6 \cdot L_x\cdot L_y \cdot L_z}$ in the thermodynamic limit.

The mapping to the square ice model can be understood as follows: for all $A$ tetrahedra, fixing a single spin also fixes the spin pair's orientation. The term \textit{spin pair} refers to the spin partner in the $x,y$-plane, so $\{S_1^z, S_2^z\}$ and $\{S_0^z, S_3^z\}$.  The opposite spin pair is, therefore, completely independent. However, the spin pair of one $A$-tetrahedron is connected via a $B$ tetrahedron to the opposite spin pair on another $A$ tetrahedron. In our numbering, if we consider a spin pair $\{S_1^z, S_2^z\}$, then this is connected via a $B$ tetrahedron to the spins $\{S_0^z, S_3^z\}$ of the $A$ tetrahedron above. So, we consider all $B$ tetrahedra with the same $z$ coordinate as a decoupled plane in the system. This can be seen equivalently as all $\{S_1^z, S_2^z\}$ of $A$ tetrahedra with common $z$ coordinates and all $\{S_0^z, S_3^z\}$ of the $A$ tetrahedron above ($z+1$), constituting a plane. 
Now we see that each $B$-tetrahedron in the plane acts as a vertex, the connecting $\{S_0^z, S_3^z\}$ in the upper plane as the vertical links and the $\{S_1^z, S_2^z\}$ in the lower plane as the horizontal links of a square lattice. An example of an $x,y$-plane of tetrahedra is shown in Fig.~\ref{fig:Lines_Planes}(a). In the Fig.~\ref{fig:Lines_Planes}(a) an $x,y$-plane of $A$ tetrahedra is shown by the blue colored tetrahedra and the lavender bonds correspond to the $B$-tetrahedra bonds. In the planar ice case, individual planes of $B$ tetrahedra decouple, so the colors of the bonds must be inverted. The blue tetrahedra in the figure now represent the $B$ tetrahedra and the lavender bonds represent the bonds of the $A$ tetrahedra. These $A$-bonds form a square lattice and fix the spins of the connecting $B$ tetrahedra that now act as vertices when projected onto the $x,y$-plane.

It is also possible to make the same Pauling estimate for this case, where we now consider that there are $\left( \frac{12}{2^7} \right)^{N/4}$ ground state configurations on an $A-B$ tetrahedron pair. This gives an estimate for the ground state degeneracy of $\Omega = (3/2)^{N/4} $, which is slightly smaller than the exact value. In a single plane, flipping a closed loop of spins aligned head to tail will, again, generate all possible ground state configurations. Importantly, on the $A$ tetrahedra, however, the spin pairs must always be flipped together; only on the $B$ tetrahedra can we freely choose a spin partner. This corresponds exactly to the loop update in the planar six-vertex model.  
When we now consider a single partial charge excitation $\mathsf{q}_A^{03/12}$ located at an $A$ tetrahedron in a certain plane, then we can see that these are bound to stay within the same plane. The partial charge can be moved by flipping the non-spin-pairs on the $A$ tetrahedron, where the charge is located, and then creating a big loop of spins aligned head-to-tail on $B$ tetrahedra, and flipping corresponding spin-pairs on $A$ tetrahedra. Lastly, to close the loop, a non-spin-pair on an $A$ tetrahedron in the same plane must be flipped, creating a partial charge. 
Monopole-excitations  $\mathsf{Q}_{A/B} = \pm2$ are also bound to the two-dimensional plane they are created in and to the tetrahedral sublattice. On the $A$ tetrahedra, they are always accompanied by a partial charge. Monopole excitations can be moved to another tetrahedron of the same kind by second-order processes, crossing $A$ tetrahedra only by the corresponding spin pair.

In summary, we have a system where the planes decouple in the ground state limit, and in each plane, the same constraints appear as in the six-vertex model. The ground state degeneracy, however, still grows exponentially with system size. Excitations are energetically bound to the same two-dimensional plane.

\subsection{Omni-Plane, $\delta_A <0,  \delta_B = 0$}
\label{Ssec:OmniPlane}
In the case $\delta_A < 0$ and $\delta_B = 0$, we also distinguish between the $A$ and $B$ sublattice. On the $A$ sublattice, the total charge must be $\mathsf{Q}_A = 0$ and the partial charge $\mathsf{q}_A^{03} = -\mathsf{q}_A^{12} = \pm 2$, while on the $B$ sublattice only the total charge must be $\mathsf{Q}_B = 0$; the value of the partial charge is not fixed. This means that 2 ground states are possible on the $A$-sublattice tetrahedra, the ones with magnetization vector parallel or antiparallel with the global $z$ axis, while on the $B$-sublattice tetrahedra, all six ground states are allowed. Starting from a ground state configuration, we can now generate all possible ground states by flipping all spins on $A$ tetrahedra in a common plane since flipping a plane of $A$ tetrahedra means also flipping two spins connected by a $B$ tetrahedron, which ensures we still stay in a ground state configuration. An example of such a plane is shown in Fig.~\ref{fig:Lines_Planes}(a). Since any plane of connected $A$ tetrahedra could be flipped, we will refer to this phase as \textit{omni-plane} phase. In each spatial direction $\alpha\in \{x, y, z\}$ there are $2 L_{\alpha}$ individual planes of $A$ tetrahedra. Along each spatial direction, we can flip up to $2L_{\alpha}$ planes at the same time. We can choose $\binom{2L_{\alpha}}{i}$ possible ways to flip $i$ planes of $A$ tetrahedra. If we count all the possible flips, we will, however, count six states twice. To account for this double-counting, we do not consider the initial and fully flipped state in all countings. So in total, there are $\sum_{\alpha =x,y,z} \sum_{i=1}^{2L_{\alpha}-1} \binom{2L_{\alpha}}{i} = \sum_{\alpha =x,y,z}(2^{2L_{\alpha}}-2)$ total ground states for this system. A more detailed description of the double counting and the projection of the ground states is given in App.~\ref{sec:App_OmniPlane}. 

A plane perpendicular to the $x$ or the $y$ direction is flippable only if the $A$ tetrahedra in this plane are ferromagnetically aligned; all $A$ tetrahedra in this plane must be in the same configuration. If this were not the case, flipping the plane would create a double monopole in adjacent $B$ tetrahedra. On the other hand, a plane perpendicular to the $z$ direction is only flippable if the $A$ tetrahedra in this plane are antiferromagnetically aligned. A flip in any plane creates a line of defects in the two perpendicular planes. 
Partial charge excitations and monopole excitations are dynamically immobile, as any crossing of an $A$ tetrahedron necessarily increases the energy of the system by $4|\delta_A|$.

In summary, we have a system where the intersecting planes partially decouple, and the ground state degeneracy grows exponentially with the linear system size or the number of planes in a spatial direction. Individual excitations on top of the ground state are fully immobile. 

\subsection{Line-Order, $\delta_A , \delta_B > 0$}
\label{Ssec:LineOrder}
In the case $\delta_A,\delta_B > 0$, both the total charge and also the partial charges are $\mathsf{Q}_{A/B} = \mathsf{q}_{A/B}^{03/12} = 0$. This means there are four possible ground states on each tetrahedron. Fixing a single spin on a tetrahedron means the other spin pair is also fixed. Since this holds on both sublattices, if a single spin is fixed -- let's say an $S_0^z$ at position $(x,y,z)$ -- also the neighboring spins ($S_3^z$ on both connecting tetrahedra) are fixed. By this principle, fixing a single spin fixes the orientation of all spins on a line -- here, all spins $S_{0/3}^z$ in $x,y$-direction connecting the initially chosen spin. A visual representation of the spin lines is shown in Fig.~\ref{fig:Lines_Planes}(b). This is the reason we refer to this phase as \textit{line-order}. All ground states can be created by flipping all spins in a spin line. 

If we have a lattice of size $\{L_x, L_y, L_z\}$, with $L_\alpha$ being the number of unit cells in direction $\alpha$, then the number of decoupled spin lines is given by the greatest common factor (GCF) of $L_x$ and $L_y$ multiplied by four times the height $L_z$ of the system. The first factor, $\text{GCF}(L_x, L_y)$, counting the number of spin-lines in a single $x,y$-plane, stems from the periodic boundary conditions, connecting spin-lines in a specific way. If we were to work with open boundary conditions, the number of spin lines in a single plane would be significantly higher. The factor $4$ is needed because in a single unit cell, we can find four decoupled spin-lines. So finally, we note that each spin-line can have two configurations independently, which gives a ground state degeneracy of $2^{4 \cdot \text{GCF}(L_x, L_y) \cdot L_z }$.
We can create four partial charges at the corners of a closed loop within a tetrahedral plane. At the four corners, the loop must pass through the tetrahedra by not flipping the corresponding spin pair and creating a partial charge. The partial charges can not be moved independently, but only at the corner of this loop, acting therefore as membrane-operators~\cite{Vijay2015Dec, Vijay2016Dec}. On the other hand, flipping a single spin creates two monopole-charges $\mathsf{Q} = \pm2$. Since they can only appear in combination with a partial charge, the monopoles can only travel along the one-dimensional spin-pair line they are created on, resembling therefore lineons~\cite{Vijay2015Dec, Vijay2016Dec}. Crossing a tetrahedron in any other direction creates additional partial charges and thus increases the energy. 

In summary, we have a system with a ground state degeneracy, which grows with the cross-section (perpendicular to the spin-pair plane) of the system or the number of spin lines parallel to the $x,y$-plane. Along the lines of spins that lie in the strain plane, the spins are aligned head to tail. Partial charge excitations are mobile in a two-dimensional plane at the corner of membrane operators, and monopole excitations are only mobile along the one-dimensional spin-pair lines.

\subsection{Plane-Paramagnetic, $\delta_A >0, \delta_B < 0$}
\label{Ssec:PlanePara}
In the case $\delta_A >0, \delta_B < 0$, we assume different signs for the two different sublattices. On both sublattices the total charge must be $\mathsf{Q}_{A/B} = 0$; on the $A$ sublattice, the partial charge $\mathsf{q}_A^{03/12} = 0$, while on the $B$ sublattice the partial charge $\mathsf{q}_B^{03} = -\mathsf{q}_B^{12} = \pm 2$. Here, 4 ground states are possible on the $A$-sublattice tetrahedra, and on the $B$-sublattice tetrahedra, 2 are allowed. The magnetization vector of the $B$ tetrahedra points along the global $z$ direction. If one spin is fixed on a $B$-sublattice tetrahedron, all spins on the whole tetrahedron are fixed, while on the $A$-sublattice, fixing one spin fixes only the opposite spin pair. Similarly to before, if we fix a single spin, we will now completely fix the neighboring $B$-tetrahedron and the opposite spin pair on the $A$-tetrahedron. Carrying this on leads to fixing a whole $x,y$-plane of $B$ tetrahedra by fixing a single spin on this plane. Since within planes, spins are ordered, but between planes, there is no correlation, we refer to this phase as \textit{plane-paramagnetic}. Within a single plane, the magnetization vectors of the $B$ tetrahedra are antiferromagnetically aligned. 
Here, a plane is given by all $B$-tetrahedra with the same $z$ coordinates, or equivalently, all the $S_{0/3}^z$ of the $A$-tetrahedra above, as well as the $S_{1/2}^z$ of the $A$-tetrahedra below. A visual representation of such a plane is shown in Fig.~\ref{fig:Lines_Planes}(a), where the blue tetrahedra represent the $B$ tetrahedra and the lavender bonds represent the bonds on the connecting $A$ tetrahedra. Since each plane can change between the two configurations independently, the ground state degeneracy is $2^{2 \cdot L_z}$, where $L_z$ is the number of unit cells in the $z$ direction. In each unit cell, there are two such planes. Naturally, all ground-state configurations can be generated by flipping planes individually. 

It is, however, important to note that with periodic boundary conditions, the number of tetrahedra of the same kind in $x$ and $y$ direction must be even, as otherwise, no ground state can be found. 
On a single line of connecting spin pairs, the pairs on the $A$ tetrahedra must be oppositely aligned, while on the $B$ tetrahedra, they must be equally aligned. So, for a system of $l=4$ $A$ tetrahedra in one direction, a possible configuration might be ($+--++--+$), where we started with an $A$ tetrahedron. The last $B$ tetrahedron closes over the boundary ($+...+$). For $l=3$, no such configuration can be found, as ($+--++-$) does not give a ground state for the last $B$ tetrahedron connected over the boundary ($+...-$). Since we are working with the convention of a single unit cell already containing an even number of tetrahedra of the same kind, we will not encounter this problem.
Partial charge and monopole excitations on top of the ground state are dynamically immobile as crossing any $B$ tetrahedron necessarily increases the energy by $4 |\delta_B|$. 

In summary, we have a system with decoupled planes, each with two possible configurations. Within a plane, neighboring $B$ tetrahedra are antiferromagnetically aligned. Individual excitations on top of the ground state are immobile by energetic constraints.

\subsection{Ferromagnetic, $\delta_A ,\delta_B < 0$}
\label{Ssec:Ferromagnetic}
In the case $\delta_A,\delta_B < 0$, the total charge is $\mathsf{Q}_{A/B} = 0$ and the partial charges are $ \mathsf{q}_{A/B}^{03} = -\mathsf{q}_{A/B}^{12} = \pm 2$. On each tetrahedron, there are two possible ground states -- the ones with magnetization vector parallel or antiparallel to the global $z$ axis.
By fixing a single spin on a single tetrahedron, however, we will fix all the spins in the lattice, as the tetrahedra are corner-sharing. So, in this case, the total ground state degeneracy is only $2$. The spin configuration on all tetrahedra of the same kind is equal. However, the magnetization vector of tetrahedra of different kinds will still point in the same direction, either into $+\hat{z}$ or $-\hat{z}$. This means that this limiting case can be seen as a ferromagnet, where the ground state spontaneously breaks the symmetry of the Hamiltonian, as pointed out in~\cite{Jaubert2010Aug}. The ground state of the system is the fully polarized state, where the local magnetization vector of all tetrahedra points in the same direction. For this reason, we refer to this phase as \textit{ferromagnetic}.
Also in this case, partial charge excitations and monopole excitations are immobile, since crossing any tetrahedron necessarily increases the energy of the system by $4 |\delta|$.

In summary, here we have a system with a double-degenerate ferromagnetic ground state. We can change the ground state by flipping all the spins of the system simultaneously. Here, the $\mathbb{Z}_2$ symmetry of the Hamiltonian is broken spontaneously.

\section{Thermodynamic properties}
\label{sec:Thermodynamics}

For any finite value of $\delta_{A/B}$, the full model's ground state degeneracy is reduced compared to the isotropic spin ice case. We can see that, depending on the respective sign of the two distortions, the ground state degeneracy can either scale exponentially with the full system volume (as in the \textit{planar ice} case) -- or with just a lower-dimensional cut of the system. When considering how to generate all possible ground state configurations within a single phase, we see that in all cases, lower-dimensional cuts of the system decouple and can be changed individually.  

\begin{figure}[h!]
    \centering
    \includegraphics[width = 1 \columnwidth]{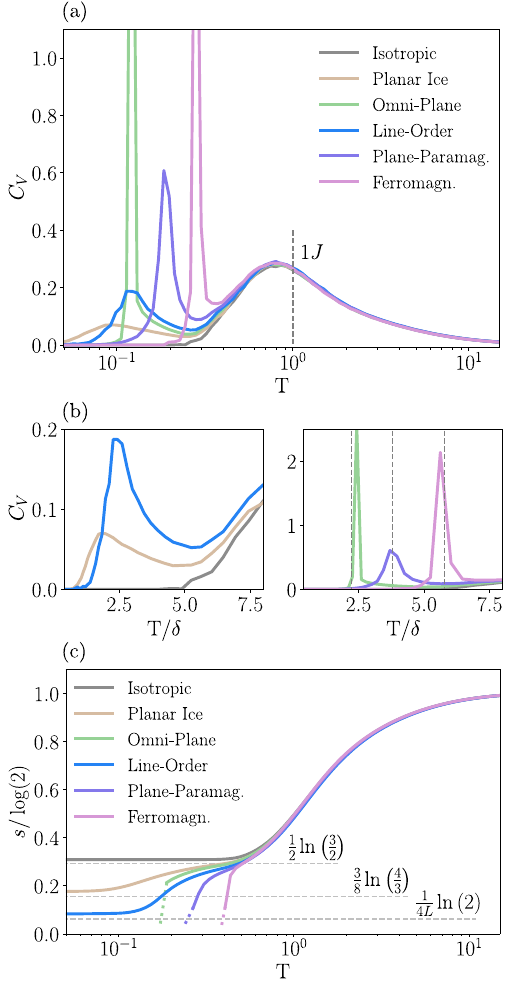}
    \caption{\textbf{Specific heat and entropy density of anisotropic spin ice.} (a) While the isotropic case shows only one bump in the specific heat, all anisotropic cases show a second bump at a lower temperature. 
    (b) Zoom in on the second bump in the specific heat. The temperature scale is linear and rescaled by $\delta$.  (c) Entropy density for all anisotropic cases. The entropy density of isotropic spin ice approaches the Pauling residual entropy value $\frac{1}{2}\ln{\left(\frac{3}{2}\right)}$, while for all other anisotropic cases, the entropy density is lowered. For planar ice the entropy density approaches the finite value $\frac{3}{8}\ln{\left(\frac{4}{3}\right)}$ and for the line-order case the entropy density approaches $\frac{1}{4L}\ln(2)$. In the symmetry-broken phases the entropy density drops to zero below the transition. All data is obtained for  $L_x = L_y = L_z = 4$ face-centered cubic (fcc) unit cells, corresponding to $N_{\text{spin}} = 1024$.   }
    \label{fig:Cv_S}
\end{figure}

Now, we will investigate how the reduced ground state degeneracy influences thermodynamic signatures.
We focus on the specific heat $C_V = \frac{\langle E^2 \rangle - \langle E \rangle^2}{T^2}$ and the entropy. By integration of the specific heat, the entropy can be calculated as a function of temperature. It is given by
\begin{equation}
    \label{eq:entropy}
    S(T) = S_{\infty} - \int_T^{\infty} dT' \frac{C_V(T')}{T'},
\end{equation}
where $S_{\infty} = N \log(2)$ is the infinite temperature entropy stemming from the $2^N$ possible spin orientations. 
These measures are indicative of the system's equilibration properties. 
Numerical data is obtained via Monte Carlo simulations, where updates are performed through attempted single spin-flips as well as worm loop updates and cluster updates to accelerate equilibration. Worm loop updates consist of finding a closed loop of spins aligned head to tail and flipping all spins part of the loop~\cite{Barkema1998Jan, Isakov2004Sep, Sandvik2006Apr}. The closed loop cannot create or destroy monopoles, but can create and move partial charges. The flip is then accepted with the Metropolis probability of $\text{min}[1, \exp{(-\Delta E/T)}]$. Since, for the anisotropic systems, the acceptance probability of a random worm loop is drastically reduced when entering into the restricted ground state manifold, additional cluster updates are employed. These cluster updates are specifically tailored to the ground state-generating operator structure. A detailed description is given in App.~\ref{Sec:App_ClusterUpdates}.
The system is first equilibrated at a specific temperature before measurements are taken. A measurement series at a specific temperature is completed when $N_{\text{sweeps}} = N_{\text{spin}} \cdot 10^3$ update-attempts are accepted. All further details of the numerical data are given in App.~\ref{sec:App_MonteCarlo}. The data presented in Fig.~\ref{fig:Cv_S} is obtained for $L_x = L_y = L_z = 4$, corresponding to $N_{\text{spin}} = 1024$ or $L_x \times L_y \times L_z = 64$ face-centered cubic (fcc) unit cells.

We will discuss specific heat and entropy for the isotropic spin ice case, $\delta_{A/B} = 0$, as well as five specific points in the phase diagram. We will touch on the generalization to all cases at the end of the section. 
To compare the characteristics of the five different cases, we fix the values of $\delta_A , \delta_B= \delta \cdot \{+1, 0, -1\}$. We will also assume that the exchange interaction on both sublattices is equally set to $J_A = J_B = 1$, and  $\delta_i < J_i$, such that the exchange interaction is still the dominant interaction. The latter means the order of excitations on top of the ground state is fixed such that a monopole excitation ($\mathsf{Q} = \pm 2$) is energetically more costly than a partial-charge change ($\Delta \mathsf{q}^{\alpha\beta} = \pm 2$). The data presented in Fig.~\ref{fig:Cv_S} is for $\delta = 0.05$.

\subsection{Isotropic case}
At high temperatures, spin ice is in a trivially disordered state as thermal fluctuations dominate. Here, the system is in a paramagnetic regime, and no correlations in the spin orientations can be found. Upon cooling, however, the system enters into the spin ice manifold. Here, each tetrahedron tends towards satisfying the ``\textit{two in - two out}" rule to locally minimize the energy. This crossover can be found, according to extensive numerical and experimental studies~\cite{Bramwell2021Oct, Ross2021Oct, Ramirez1999May, Bramwell2001Jul, Bramwell2001, Harris1997Sep} at around $T_c \approx J$. Above this crossover, thermal fluctuations easily introduce new monopoles. This crossover can be experimentally observed by a Schottky-like peak in the specific heat. With simple considerations for a single tetrahedron, neglecting double monopoles, we can confirm this estimate for the point of this crossover: there are six possibilities for a ground state on a tetrahedron, while there are eight configurations with a single monopole. Calculating the maximum of the specific heat obtained by the derivative of $Z = 6 + 8\cdot e^{-2\beta J}$ gives $T_c \approx 0.802 J$. Following eq. \eqref{eq:entropy}, also, the entropy at a specific temperature can be computed. It can be observed that for temperatures below this crossover regime, the entropy density approaches a finite value of $s_{\text{res}} = S/(k_B N) \approx \frac{1}{2} \ln\left( \frac{3}{2} \right)$, the Pauling residual entropy~\cite{Pauling1935Dec}. This has been reported in experiments as well, for example, in Dy$_2$Ti$_2$O$_7$~\cite{Ramirez1999May}, indicating the large degeneracy of ground states. In Fig.~\ref{fig:Cv_S}, the isotropic spin ice is the grey curve that follows the expected behavior well.

\subsection{Anisotropic Cases}
\label{Sec:Cv_anisotropic}

The specific heat and entropy in the temperature regime from $T=0.08J-15.0J$ for the five different anisotropic phases are shown in Fig.~\ref{fig:Cv_S}.
For all of our analyzed cases, we can observe that the specific heat follows the behavior of spin ice until the crossover at $T \approx J$. This can be understood as all the systems having similar behavior at high temperatures: a paramagnetic regime where thermal fluctuations govern the system. 
At around the first crossover, the behavior of the isotropic and the anisotropic systems still matches. Here, the system crosses over from a paramagnetic regime to a regime where spins on a tetrahedron follow the ice rule. However, below this point, the behavior of the different phases is very different. In fact, we can observe either a second crossover (in the \textit{planar ice} phase), indicated by a second bump in the specific heat, or a phase transition toward an ordered phase, indicated by a divergence in the specific heat. This allows us to make a distinction between the symmetry-broken phases and the non-symmetry-broken phases. 
For all systems, we can observe that the entropy density is lower than the Pauling estimate. 
Next, we analyze the five different phases in more detail and distinguish between the non-symmetry-broken phases and the symmetry-broken phases at finite temperatures.

\subsubsection{Non-symmetry-broken phases}
We can identify two phases that do not spontaneously break the global $\mathbb{Z}_2$ symmetry or any intermediate symmetry of the Hamiltonian at any finite temperature: the \textit{planar ice} and the \textit{line-order} phase.

For the \textit{planar ice} case, we consider only $\delta_A > 0$. Here, a second wide peak in the specific heat can be found at around $T_{\delta_A} \approx 2 \delta_A$ that indicates the crossover into the restricted ground state manifold. We can also observe that the residual entropy density does not approach the value indicative of spin ice, $s_{\text{res}} = \frac{1}{2} \ln\left( \frac{3}{2} \right) = 0.203$, but a lower value is reached. In accordance with the ground state degeneracy, the value $s_{\text{res}} = \frac{3}{8} \ln\left( \frac{4}{3} \right) = 0.108$ is approached. Here, even though the entropy density is lowered, the residual entropy density is still finite.

For the \textit{line-order} case, we observe a broad peak in the specific heat that shifts to lower temperatures as the system size increases. This behavior is consistent with an effective dimensional reduction and a mapping to a one-dimensional Ising chain with antiferromagnetic interactions at low temperatures. As expected, akin to a 1D Ising system, no finite-temperature phase transition occurs~\cite{Batista2005Jul}; however, at $T=0$, the ground state is indeed symmetry broken. We expect the specific heat of the model to show a Shottky-like peak around $T_{\delta}\approx 0.8 \cdot 2\delta$ in the thermodynamic limit.

If, the $\delta$-parameters on the two sublattices were different, $\delta_A \neq\delta_B >0$, then we will observe a first peak around $T_{\delta_A} \approx 2 \delta_A$ and a second peak around $T_{\delta_B} \approx 2 \delta_B$. If the two are equal, the peaks approximately stack in their common crossover regime.
We can again see the entropy density drop below the Pauling estimate, approaching zero for increasing system sizes as $s_{\text{res}} = \frac{1}{4 \cdot L_xL_y} \ln{(2^{\text{GCF}(L_x, L_y)})}$.

\subsubsection{Symmetry-broken phases}
If on any sublattice the strain-parameter $\delta_A/\delta_B <0$, a phase transition into a symmetry-broken state will occur at a finite temperature. 
The phase transition presents itself as a divergence in the specific heat. Due to finite-size effects, the divergence will appear as a narrow peak with a finite height in Fig.~\ref{fig:Cv_S}.

In the \textit{omni-plane} case, $\delta_A<0, \delta_B=0$, the spins on the $A$ tetrahedra will order in such a manner that the local magnetization vector of the tetrahedron is $M\parallel \hat{z}$. We can now distinguish between three types of ground states: the $A$-tetrahedra will either order antiferromagnetically in all individual planes perpendicular to the $z$ direction, or they will order ferromagnetically in all individual planes perpendicular to the $x$ or $y$ direction. This is shown graphically in App.~\ref{sec:App_OmniPlane}.  
In all cases, the $\mathbb{Z}_2$ symmetry in the respective plane is broken, and therefore, a phase transition will appear. The critical temperature is approximately $T_c \approx\frac{4\delta}{\log{(6)}}$. A discussion on the critical temperature is given in App.~\ref{sec:App_OmniPlane}. Due to the phase transition, the entropy drops sharply below the critical temperature. Since numerical integration of the finite-sized peak can lead to large inaccuracies, the drop is only indicated by a dotted continuation of the data points. 

In the \textit{plane-paramegntic} case, $\delta_A>0$ and $\delta_B<0$, the local magnetization vector of all $B$ tetrahedra are $\mathbf{M}_i \parallel \hat{z}$. Within single $x,y$-planes of $B$ tetrahedra, the $B$ tetrahedra form a square lattice. Due to the order of the spin-pairs on $A$-tetrahedra, the $B$ tetrahedra will order antiferromagnetically within the planes and break the planes $\mathbb{Z}_2$ symmetry. This transition to antiferromagnetic order in the decoupled planes will occur at a critical temperature of $T_c = \frac{4 \delta}{\log{(1+\sqrt{2})}}$~\cite{Onsager1944Feb}, the same critical temperature expected in a square aniferromagnetic Ising model. At the critical point, the specific heat is expected to diverge, and a steep drop in entropy is visible. The drop is, again, indicated by the dotted continuation of the data points.

For the \textit{ferromagnetic} case, where $\delta_A = \delta_B <0$, it has been shown that a transition akin to the one found in hydrogen-bonded ferroelectrics known as the KDP (potassium dihydrogen phosphate) model~\cite{Lieb1967Jul, Slater1941Jan} takes place, ordering the system ferromagnetically. Here, a sharp divergence of the specific heat, as well as a sharp dip in the entropy, is visible~\cite{Jaubert2010Aug, Jaubert2017May}. We can calculate the exact critical temperature by considering the energy cost of a string excitation and the entropy of this excitation~\cite{Jaubert2008Feb}. Let us consider a ground state of the system. All the spins are equally aligned on all tetrahedra, such that the global magnetization vector of all tetrahedra points in the same direction. A string excitation, now, consists of flipping a row of spins in the $z$ direction and thereby increasing the energy. Since we are introducing a partial charge in 4 tetrahedra per unit cell, for $L_z$ unit cells in $z$ direction, the energy for such an excitation is $\Delta U = 4\delta \cdot 4 L_z = 16 \delta L_z$. Because there are two possible ways for a string to enter each tetrahedron, and we flip $4L_z$ tetrahedra, the entropic gain of a string excitation is $\Delta S = \ln(2^{4 L_z})$. So the difference in free energy for a string excitation is 
\begin{equation}
    \Delta F = \Delta U -  \Delta S = 4 L_z ( 4 \delta - T \ln(2)). 
\end{equation}
This gives us a critical temperature of $T_{c,\delta} = \frac{4\delta}{\ln(2)}$, agreeing with our numerical data. For the case, $\delta_A \neq \delta_B$, the system will first order into the \textit{omni-plane} order at $T_{\delta_A}\approx\delta_A$ where we assume $|\delta_A| > |\delta_B|$ before it will order ferromagnetically at $T_{\delta_B} \approx \delta_B$.

In summary, in all cases, the specific heat has a second signature below the first crossover at $T\approx J$. If on any sublattice there is a negative strain parameter, $\delta_i <0$, we expect a phase transition at a finite temperature into a symmetry-broken phase. The symmetry breaking will, however, occur only in dimensionally reduced subsystems if one of the strain parameters is non-negative. Here, the order parameter is defined on a dimensionally reduced support -- a two-dimensional plane -- and the symmetry that is broken is the intermediate symmetry of the plane~\cite{Nussinov2009Oct, Nussinov2012Oct, Nussinov2015Jan}. This also leads to a vanishing entropy density.
If both strain parameters are positive, one-dimensional lines will effectively decouple, leading to antiferromagnetic order on lines at zero temperature. Similar behavior has been observed in the anisotropic antiferromagnetic triangular lattice Ising model, where strong coupling along one lattice direction leads to effective dimensional reduction and frustrated inter-chain interactions. Despite the two-dimensional connectivity, the system may exhibit behavior akin to weakly coupled one-dimensional chains without a finite-temperature phase transition~\cite{Houtappel1950May}. If only one strain parameter is non-zero and positive, a crossover into a regime where the two-dimensional planes decouple with an extensive ground state degeneracy occurs. Within planes, the \textit{planar ice} constraint holds. This is the only phase where the entropy density stays finite even for an infinite system size. 

Lastly, it is important to note that since the second ordering or transition into the restricted ice manifold takes place around roughly $T_{\delta}\approx |\delta_i|$ to $6 |\delta_i|$, the dip in the entropy can only be visible at temperatures in this regime. For small values of $\delta$, there is, in fact, a large temperature regime where the entropy seems to perfectly approach the spin ice estimate even for the anisotropic systems.

Here, we assume the systems to be at equilibrium. To reach thermal equilibrium in numerical simulations, techniques such as cluster updates or worm loop updates are employed to overcome large energy barriers. In fact, we can observe that the acceptance rate for single spin-flips is reduced significantly below the first crossover into the spin ice regime at $T \approx J$ (see App.~\ref{Sec:App_ClusterUpdates}). For isotropic spin ice, the worm algorithm has been proven useful to overcome the long waiting times to overcome the energy barriers and equilibrate spin ice systems~\cite{Melko2004Oct, Barkema1998Jan, Isakov2004Sep, Sandvik2006Apr}. The acceptance rate for worm loop updates is exactly one below the crossover into the spin ice manifold for the isotropic case. However, for the anisotropic systems, even the acceptance rate for the worm algorithm reduces close to the crossover into the respective restricted manifold. This renders equilibrating the anisotropic systems numerically very hard, as updates are correlated with extremely long waiting times that increase with increasing system size. It is, however, important to point out that these are higher-order processes that will also take extremely long in an experimental setup. It is expected that cooling the system too quickly will hide these features, and the system might then seem to behave akin to spin ice~\cite{Pomaranski2013Jun}.

\section{Spin Structure Factor}
\label{sec:SSF}

Neutron scattering experiments for spin ice materials uncover particular features of the ground-states~\cite{Bramwell2001Jul,Fennell2009Sep, Isakov2004Oct, Conlon2010Jun}. Since the ground state structure changes in the anisotropic models, the static spin structure factor is expected to look different.
In polarized neutron scattering experiments, one can distinguish between the spin-flip (SF) channel and the non-spin-flip (NSF) channel. The SF channel is sensitive to the spin component orthogonal to the neutron polarization, which induces a flipping of the neutron moment, while the NSF channel is sensitive to the spin components parallel to the neutron polarization. 

In a scattering event with neutron polarization parallel to $\hat{z}_{\text{sc}}$, one can define the orthonormal basis vectors for each scattering wave vector $\mathbf{q}\perp \hat{z}_{\text{sc}}$ as $\hat{x}_{\text{sc}} \equiv \hat{\mathbf{q}}$ and $\hat{y}_{\text{sc}} \equiv \hat{z}_{\text{sc}} \times \hat{x}_{\text{sc}}$. The spin vector is defined as $\hat{S}^z_{i,\alpha} =S^z_{i,\alpha}\cdot \hat{e}_i $, where $\alpha$ is the index referring to the four sites in the face-centered cubic (fcc) unit cell, while $i$ is the spin index, referring to the site on a single tetrahedron. The unit vectors $\hat{e}_i$ are the local basis vectors pointing into the center of a single tetrahedron, as defined in App.~\ref{Sec:App_LocalSpinBasis}.

Now, the respective spin structure factor channels are proportional to the following:

\begin{equation}
\label{eq:SF}
    S_{\text{SF}}(\mathbf{q}) = \sum_{i,j} (\hat{z}_i\cdot \hat{y}_{\text{sc}}) \langle S_i^*(\mathbf{q})S_j(\mathbf{q}) \rangle (\hat{z}_j\cdot \hat{y}_{\text{sc}}),
\end{equation}
and 
\begin{equation}
\label{eq:NSF}
    S_{\text{NSF}}(\mathbf{q}) = \sum_{i,j} (\hat{z}_i\cdot \hat{z}_{\text{sc}}) \langle S_i^*(\mathbf{q})S_j(\mathbf{q}) \rangle (\hat{z}_j\cdot \hat{z}_{\text{sc}}).
\end{equation}
Here, we neglected to consider the electric form factor of the magnetic ions, as it is a numerical prefactor that is not relevant to our calculations~\cite{Lovesey1986Oct, Chung2022Mar}.

In polarized neutron scattering experiments, with $\hat{z}_{\text{sc}} \parallel [1\bar{1}0] $, the spin ice material Ho$_2$Ti$_2$O$_7$ and Dy$_2$Ti$_2$O$_7$ show particular pinch points in the $[hhl]$ plane in the SF channel, indicative of the dipolar form of the spin correlations in three dimensions, while in the NSF channel no particular features can be seen~\cite{Bramwell1998Apr, Fennell2009Sep}. 
Now, we want to understand how this structure factor will differ in the anisotropic phases compared to isotropic spin ice. Numerically, the spin structure factor can be obtained by sampling states via a Monte Carlo procedure or by an analytical self-consistent Gaussian approximation. We will employ both techniques and show the correspondence of the data in Fig.~\ref{fig:SSF_SF}.

\subsection{Monte Carlo sampling}
\label{Ssec:MC_sampling_SSF}
We collect states by Monte Carlo sampling a system of size $L=10$ (corresponding to $N_{\text{spins}} =16000 $ spins) and compute the neutron scattering response. For $T\rightarrow0$ states can be directly computed by sampling within the ground state manifold. To this end, sampling algorithms that allow us to generate all possible ground states of an anisotropic phase are employed. These sampling algorithms have a transition probability of $1$ when starting from a state within the ground state manifold and are described in detail in App.~\ref{Sec:App_ClusterUpdates}. 

For the \textit{isotropic} spin ice case, we sample the ground states with a worm algorithm, where we find a closed loop of spins aligned head to tail and flip them all. Starting with a ground state configuration, this will always bring us back to a ground state.
In the \textit{planar ice} case, a similar algorithm can be employed; however, here, the loops only lie in the ice planes - or the plane spanned by a layer of $B$ tetrahedra with the same $z$ coordinates.
In the \textit{omni-plane} case, we randomly choose an axis and flip a random plane of $A$ tetrahedra perpendicular to this axis. We check that after the flip, the state is still in the ground state manifold; otherwise, discard the state.
For the \textit{line-order} case, we can again start with a ground state and sample other ground states by randomly flipping lines of connected spin pairs spanning over the system's boundaries.
Similarly, in the \textit{plane-paramagnetic} case, ground states can be sampled by randomly flipping planes of $B$ tetrahedra with equal $z$ coordinates.
The \textit{ferromagnetic} case contains only two ground states that can be easily constructed numerically for any system size by putting the same ground state on all tetrahedra.

\subsection{Self-Consisteng Gaussian Approximation}
\label{Ssec:MFT_SSF}
It has been confirmed by comparison to Monte Carlo data that the correlations of spins at low temperatures of the nearest-neighbor \textit{isotropic} spin ice Hamiltonian are accurately described by the approximation of an \textit{N}-component spin~\cite{Reimers1991Jan, Garanin1999Jan, Canals2002Apr, Isakov2004Oct, McClarty2014May} and stay correct even when including further-neighboring terms or long-range interactions~\cite{Conlon2010Jun, Enjalran2004Nov}.
With the help of a self-consistent Gaussian approximation (SCGA), the spin-spin correlations, or the spin structure factor, can also be computed analytically in the limit of $N \rightarrow \infty$~\cite{Conlon2010Jun}
\begin{equation}
\label{eq:Corr_MFT}
    S_{\alpha \beta}(\mathbf{q}) =  \left[ \lambda \delta_{\alpha, \beta} + \beta J_{\alpha \beta}(\mathbf{q}) \right]^{-1},
\end{equation}
where the Lagrange multiplier $\lambda$ fixes the spin length $\braket{S_{\alpha}^2} = 1$ via
\begin{equation}
\label{eq:SelfConsistentEq}
    \braket{S_{\alpha}^2} = \frac{1}{4 L^3} \sum_{\mathbf{q} \in \text{BZ}} \text{Tr}\left[ \lambda \mathbb{I} + \beta \hat{J}(\mathbf{q}) \right]^{-1}.
\end{equation}
Here, $L$ is the linear system size, and we insert the Fourier transform of our interaction Hamiltonian, specified in App.~\ref{sec:App_MFT}. The Lagrange multiplier $\lambda$ can be interpreted as the thermal occupation value of the lowest energy mode~\cite{Chung2022Mar} and will drop to zero if there is a phase transition in SCGA~\cite{Chung2022Mar, Conlon2010Jun}. An analysis of the Lagrange multiplier in the different $\delta$-regimes can be found in App.~\ref{sec:App_MFT} and is consistent with the findings of section~\ref{sec:Thermodynamics}.

The scattering result in the SF channel in the SCGA is computed by inserting the correlator Eq.~\eqref{eq:Corr_MFT} into Eq.~\eqref{eq:SF}. To resolve this at different temperatures, we first need to solve for $\lambda$ in \eqref{eq:SelfConsistentEq} at the necessary inverse temperature $\beta$. 
We can also compare the result from the SCGA to numerical data, where we sample ground states for each system according to the ground state rule. In the SCGA, we compute the correlations at a temperature well below the crossover ($T \ll 1J$) to accurately reproduce correlations close to or at the ground state.

\subsection{Results}
\begin{figure*}
    \centering
    \includegraphics[width = 1. \linewidth]{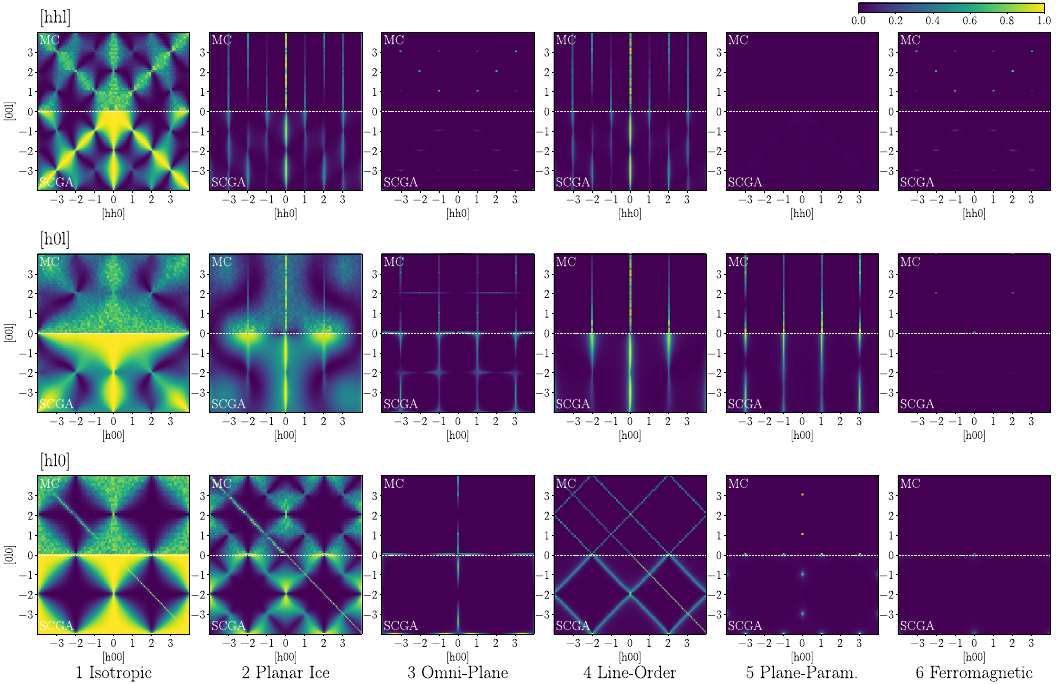}
    \caption{\textbf{Normalized spin structure factor in Spin-Flip (SF) channel.} Columns show the same system, and rows have the same cut in the Brillouin zone. \textit{Columns :} 1 Isotropic, 2 Plane Ice, 3 Omni-Plane, 4 Line-Order, 5 Plane-Paramagnetic, 6 Ferromagnetic. \textit{Rows:} 1 [hhl], 2 [hl0], 3 [h0l]. Every panel shows Monte Carlo (MC) data on the \textit{top} and self-consistent Gaussian approximation (SCGA) on the \textit{bottom}. The Monte Carlo data is obtained by sampling in the degenerate ground-state sector for systems with $L_x = L_y = L_z = 10$ fcc unit cells, corresponding to $N_{\text{spin}} = 16000$.    The intensities are normalized across a single system between $0$ and $1$ in a.u..}    \label{fig:SSF_SF}
\end{figure*}

The normalized SF channel of the neutron scattering cross section for the six different phases at or close to the ground state is shown in Fig.~\ref{fig:SSF_SF}. The intensity is normalized for each system, $(S_{\text{SF}}(\mathbf{q})-\text{min}(S_{\text{SF}}(\mathbf{q})))/(\text{max}(S_{\text{SF}}(\mathbf{q}))-\text{min}(S_{\text{SF}}(\mathbf{q})))$, such that the signatures will lie between $0$ and $1$. We can see that for any limit of $\delta_A, \delta_B$, the spin structure factor computed numerically shows no significant difference from the result obtained by the SCGA when the temperature is set close to the transition temperature. It is important to note that the signatures on the antidiagonal in the [hl0] cut are due to the neutron polarization being almost parallel to the scattering wave vector and not a signature of the underlying spin configuration.

In the \textit{isotropic} case, we can see the characteristic pinch-point singularities in the SF channel data. These points in the scattering data result from the spin correlations being of dipolar form, $C_{\alpha \beta}(\mathbf{r}) \propto \frac{1}{r^5} (r^2\delta_{\alpha \beta} - 3r_{\alpha} r_{\beta})$, which lead to a structure factor $S_{\alpha \beta} \propto (\delta_{\alpha \beta} - \frac{k_{\alpha} k_{\beta}}{k^2}) $~\cite{Fennell2009Sep, Youngblood1980Jun, Isakov2004Oct}. Pinch points are accompanied by ``bow-tie" correlations around point singularities located at the high symmetry points of the extended Brillouin zone of the pyrochlore lattice at positions $(111)$, $(200)$, $(220)$, and $(222)$ as well as symmetry-related points. 

In the \textit{planar ice} phase, the pinch points with bow-tie patterns are no longer present in the [hhl] and [h0l] plane. Only in the [hl0] cut do pinch points still exist with a bow-tie pattern. This is understood from the cut being exactly parallel to the ice plane. Here, in the plane, correlations are still of dipolar form, with $C_{\alpha \beta}(\mathbf{r}) \propto \frac{1}{r^4} (r^2\delta_{\alpha \beta} - 3r_{\alpha} r_{\beta})$, for $r$ in the [hl0] plane. This will lead to a structure factor of the form $S_{\alpha \beta} \propto (\delta_{\alpha \beta} - \frac{k_{\alpha} k_{\beta}}{k^2})$~\cite{Youngblood1980Jun} in the plane, due to the square ice constraint in the ground state. The location of the pinch points in square ice are located at the high symmetry points of the square lattice, $(\pm 1,0)$, $(0,\pm 1)$, and $(\pm 1, \pm 1)$. This translates to pinch points in the [hl0] cut at $(\pm100)$, $(0\pm10)$, and $(\pm1\pm10)$ and symmetry equivalent points in the extended Brilloin zone. Since single planes are not correlated, other cuts in the lattice do not show this behavior.

In the \textit{omni-plane} case, spins order in particular patterns within the different tetrahedra-planes of the lattice. A single $A$ tetrahedra has a local magnetization vector that is either parallel or antiparallel to the $z$ axis. These tetrahedra will align predominantly antiferromagnetically along planes perpendicular to the $z$ direction and predominantly ferromagnetically for planes perpendicular to the $x$ and $y$ directions. A detailed description and graphical representation of this can be found in App.~\ref{sec:App_OmniPlane}.

In a unit cell, there are six possible spin configurations. Two of which are fully ferromagnetic, leading to sharp Bragg peaks in the Brillouin zone center at $(000)$. The other four have all the $A$ tetrahedra in either the $x,z$ or the $y,z$-plane ferromagnetically aligned, while in all other planes, the $A$ tetrahedra are antiferromagnetically aligned. For these four states, however, two adjacent ferromagnetic layers have opposite polarizations. This is necessary to get the antiferromagnetic alignment in the other layers. When we now project the two ferromagnetic layers of a unit cell down onto the plane, the two oppositely polarized planes form an antiferromagnetic square lattice. For this order, characteristic Bragg peaks at the four corners of the square lattice can be observed. On the other hand, projecting the antiferromagnetic layers of a unit cell onto the plane gives a square lattice with stripe order. For a square lattice with horizontal or vertical stripe order, Bragg peaks at $(0, \pm 1)$ or $(\pm 1, 0)$ appear. 

Any plane flip in the system will introduce line defects in the square lattice and thereby broaden the sharp Bragg peaks along the flipping direction. This explains the square grid structure in the [h0l] plane; the AFM square lattice peaks are broadened by introducing line defects. In the [hl0] plane, on the other hand, the stripe order peaks at $(\pm1\pm10)$ are broadened due to the introduction of line defects and form the cross-patterns. A detailed explanation with schematic figures can be found in App.~\ref{sec:App_OmniPlane}.

In the \textit{line-order} case, spins will order along spin-lines in the $x,y$-plane. The pinch-point singularities of the isotropic model are no longer visible, but line-like patterns can be observed in the different cuts within the system. The spin lines run perpendicular to the [hhl] cut and the [h0l] cut, giving rise to a vertically striped pattern. The [hl0] cut is parallel to the spin lines; here, we can see a cross pattern being formed. 

In the \textit{plane-paramagnetic} case, spins form decoupled planes, where within planes, the local magnetization vector of the $B$ tetrahedra align antiferromagnetically. This order in the $x,y$-planes is visible as Bragg peaks in the [hl0] cut, where the $B$ tetrahedra form a square lattice. In a square lattice with AFM order, we expect Bragg-peaks to form at $(\pm 1, \pm 1)$; however, the square lattice formed by the $B$ tetrahedra of a unit cell is rotated by $\pi/2$ compared to the $x, y$-projection of the fcc unit cell. This leads to the peaks at the expected locations of $(\pm100)$ and $(0\pm1 0 )$ as well as symmetry equivalent positions in the extended Brillouin zone. The uncorrelated planes can be inferred by the vertical lines in the [h0l] cut.

In the \textit{ferromagnetic} case, the symmetry of the system is spontaneously broken, and the spins order ferromagnetically. For ferromagnetic order in the fcc lattice, Bragg peaks become visible at all points [hkl] where either all $h,k,l$ are even, or all $h,k,l$ are odd. So here, Bragg peaks are visible at $(000)$, $(200)$, $(220)$, $(222)$, $(111)$, and all symmetry equivalent points.

In all anisotropic cases, we can see that the neutron scattering data in the SF channel at low temperatures will deviate significantly from the expected result of the isotropic model. Pinch-point singularities are only visible in the \textit{planar ice} phase within the ice plane, as dipolar correlations are still present here. In all other phases, a certain order appears for the specific cuts. The cut in the [0hl] plane will look equivalent to the [h0l] cut.

\begin{figure}[h!]
    \centering
    \includegraphics[width = 1. \columnwidth]{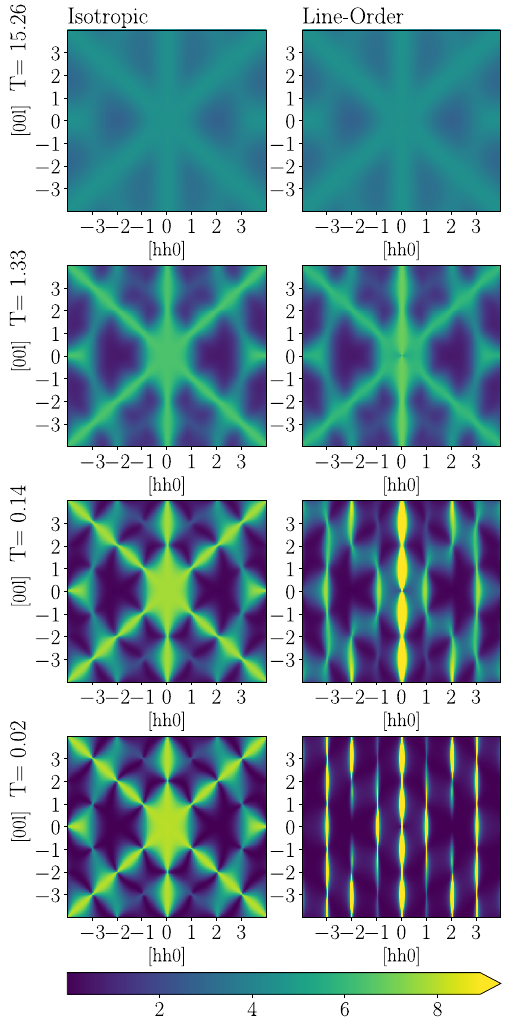}
    \caption{\textbf{SCGA spin structure factor in Spin-Flip (SF) channel at different temperatures.} While for high temperatures ($T\gg 1$) the signatures for the isotropic (left) and anisotropic (right) systems agree, when the temperature is lowered (top to bottom), the features look significantly different. The line-order regime is evaluated with $\delta_A = \delta_B = 0.05$ and the [hhl] cut in the Brillouin zone is shown. The scale is chosen uniformly across all temperatures and both systems.}    \label{fig:T_EVOLVE}
\end{figure}

While these signatures will be visible at low temperatures, very close to the ground state, remnants of the change in correlations will also be visible at higher temperatures. We show how the neutron scattering response resulting from the SCGA changes between isotropic spin ice and the anisotropic systems for decreasing temperatures with the example of the line-order case in Fig.~\ref{fig:T_EVOLVE}. Here, the colorscale is chosen uniformly across all temperatures and both systems for better comparison. We can observe that at high temperatures, for $T\gg 1J$, the scattering data will be indistinguishable between the isotropic and the anisotropic cases. However, significant differences can be observed below the spin-ice crossover, where the anisotropic systems gradually enter their respective ground-state manifold and the aforementioned signatures become visible. Measuring the structure factor as a function of temperature thus directly unveils the anisotropic couplings of the model Eq.~\eqref{Htot}.

\section{Summary and Outlook}
\label{sec:Summary}

In this work, we explored how anisotropic perturbations in the nearest-neighbor spin ice Hamiltonian affect the ground state structure and equilibrium properties of the spin system on the breathing pyrochlore lattice. By tuning the strain parameter, $\delta_{A/B}$, independently on the two tetrahedral sublattices, we demonstrated that the ground state degeneracy exhibits a sign-dependent change, leading to an additional five distinct phases. In all cases, the degeneracy is reduced; strikingly, in only one of the five anisotropic phases, the degeneracy grows exponentially with the full system volume, while in the other scenarios, the ground state degeneracy scales with a lower-dimensional cut of the system.
Additionally, we analyzed the mobility of excitations on top of the ground state. We found that any negative $\delta$ rendered excitations immobile, while a positive $\delta$ made their mobility restricted to lower-dimensional submannifolds. These behaviors mirror key features of fracton systems, where excitations are immobile or exhibit constrained dynamics due to conservation laws or energetic barriers. In our case, the excitations are energetically restricted to planes or lines, or are completely frozen, making them fracton-like in nature.

We also investigated the equilibrium properties of the system via the specific heat and entropy. Our results show that the anisotropic perturbation will either lead to a phase transition into a symmetry-broken phase at a critical temperature (\textit{omni-plane}, \textit{plane paramagnetic}, and \textit{ferromagnetic} phase), a phase transition at zero temperature (\textit{line-order}), or a crossover into a phase with decoupled planes (\textit{planar ice}). In all cases but one, the \textit{planar ice} phase, the residual entropy density approaches zero for increasing system sizes. The static spin structure factors of the anisotropic systems display significant deviations from the conventional spin ice case, showing only dipolar correlations in the \textit{planar ice} phase.

Our analysis introduces a new perspective on frustrated magnetic systems, particularly in the context of strain engineering. While strain effects in spin ice systems have been previously studied~\cite{Bovo2019Mar, Bovo2014Mar, Barry2019Aug, Jaubert2010Aug}, introducing sublattice-dependent anisotropy enriches this field by opening a qualitatively different route to dimensional reduction. A well-known example of such reduction is the kagome ice state, realized under an external magnetic field along the global $[111]$ direction~\cite{Matsuhira2002Jul, Higashinaka2003Jul, Kao2016May, Fukazawa2002Jan, Sakakibara2003May}. In that case, spins aligned with the field are pinned while the remaining spins fluctuate within decoupled two-dimensional kagome planes. By contrast, in our model, the reduction arises not from field-induced polarization but from frustration-lifting via sublattice-dependent bond anisotropy, providing an alternative mechanism for emergent lower-dimensional dynamics in spin ice.

Looking forward, several intriguing questions arise from our findings. Since the ground-state degeneracy is significantly reduced in some cases, an important question concerns the mechanisms by which equilibrium is achieved. Furthermore, the timescales on which equilibrium is approached are to be investigated. Additionally, we analyzed the mobility of excitations on top of the ground states, which made a connection to fractons apparent~\cite{Chamon2005Jan, Haah2011Apr, Yoshida2013Sep, Bravyi2013Nov, Vijay2015Dec, Yan2020Mar, Han2022Jun}. 
This raises the question of how these energetic constraints and subdimensional mobility influence the relaxation dynamics of the system.

Finally, in this study, we restricted our analysis to the nearest-neighbor Ising Hamiltonian on the pyrochlore lattice, neglecting the long-range dipolar interactions that are intrinsic to real spin ice materials~\cite{Siddharthan1999Aug, Melko2001Jul, Gingras2011Feb, Melko2004Oct}. These long-range interactions are known to induce a transition into a long-range ordered state at temperatures around $T_c \approx 0.07 D_{nn}$, where $D_{nn}$ is the nearest-neighbor dipolar interaction strength~\cite{Melko2001Jul, Melko2004Oct}. The dipolar spin ice model exhibits a set of twelve symmetry-related ground states, each characterized by a sixteen-site magnetic unit cell.

While our anisotropic model neglects these long-range terms, it still selects a smaller subset of states from the spin ice manifold, effectively reducing the degeneracy. Interestingly, we find that the ground-state construction rules of the anisotropic model generate states that, in most cases, are also ground states of the dipolar model; except in the ferromagnetic regime ($\delta_A, \delta_B < 0$), where full alignment conflicts with the long-range dipolar correlations. This indicates that there exists a class of configurations that are simultaneously ground states of both the classical spin ice model with long-range interactions and the anisotropic nearest-neighbor model with bond-dependent couplings.

This overlap suggests that introducing sublattice-resolved strain, even in the presence of dipolar interactions, could help lift the twelve-fold degeneracy of the dipolar model and potentially enhance the ordering temperature, providing a promising route toward controlled stabilization of specific spin ice ground states.

\textbf{Acknowledgements.---}
We thank Chris Laumann for insightful discussions. 
We acknowledge support from the Deutsche Forschungsgemeinschaft (DFG, German Research Foundation) under Germany’s Excellence Strategy–EXC–2111–390814868, TRR 360 – 492547816 and DFG grants No. KN1254/1-2, KN1254/2-1, FOR 5522
(project-id 499180199), the European Research Council (ERC) under the European Union’s Horizon 2020 research and innovation programme (grant agreement No 851161), the European Union (grant agreement No 101169765), as well as the Munich Quantum Valley, which is supported by the Bavarian state government with funds from the Hightech Agenda Bayern Plus.

\textbf{Data and Informations availability.---} Data, data analysis, and simulation codes are available upon reasonable request on Zenodo~\cite{zenodo}.

\appendix
\section{Experimental considerations}
\label{sec:App_Experimental_considerations}
We now discuss possible realizations of the presented toy model. The system Hamiltonian~\eqref{eq:Ham_Q} is based on the classical spin ice model. Due to large crystal field anisotropies, the spins are constrained to point along the local $\langle 111 \rangle$ directions connecting each corner point of the tetrahedron to its center. This reduces the full spin Hamiltonian to an Ising-like interaction between the spins on each tetrahedron~\cite{Bramwell1998Apr, Harris1997Sep}.
Importantly, we consider a minimal model containing only the nearest-neighbor exchange interaction. In rare-earth spin ice compounds such as Dy$_2$Ti$_2$O$_7$ and Ho$_2$Ti$_2$O$_7$, magnetic interactions include both a short-range antiferromagnetic exchange $J_{\text{nn}}$ and a long-range ferromagnetic dipolar term $D_{\text{nn}}$. These combine into an effective nearest-neighbor coupling $J_{\text{eff}} = J_{\text{nn}} + D_{\text{nn}}$, which stabilizes the spin ice regime at intermediate temperatures. However, in our theoretical approach, we adopt a minimal description by retaining only the nearest-neighbor interactions and neglecting long-range dipolar terms beyond first neighbors. This approximation captures the correct thermodynamics at higher temperatures but omits the low-temperature ordering transition seen in full dipolar spin ice, which occurs around $T_c \approx 0.077 D_{\text{nn}}$~\cite{Melko2001Jul}. Despite this simplification, the minimal model remains a valuable tool for exploring the role of anisotropy and frustration in modifying the spin ice ground-state manifold.

In this minimal model, the ground-state condition is that the net spin sum on each tetrahedron vanishes. However, this simple ground-state behavior in spin ice can change under the influence of strain, pressure, or when an external field is applied~\cite{Mito2007Mar, Edberg2019Oct, Edberg2020Nov, Jaubert2010Aug, Bovo2014Mar}. Pressure along certain crystal axes may reduce the distance between the crystal sites along this direction, causing the ionic wave function to overlap and the exchange interaction between the neighbors to change~\cite{Edberg2019Oct}, such that $H_{\mathrm{ strain}} = \sum_{\braket{ij}} J_{ij} S_i^z S_j^z$. While the precise relationship between strain and the individual contributions $J_{\text{nn}}$ and $D_{\text{nn}}$ is material dependent and beyond the scope of our minimal model, it is known that dipolar interactions ($D \sim r^{-3}$) are significantly less sensitive to lattice distortions than exchange interactions ($J \sim r^{-q}$, with $q \gg 3$)~\cite{Bramwell1990Sep, DeJongh1975Jan}. Therefore, strain primarily modulates the exchange term.

Experimental support for this assumption comes from studies on single crystals of Dy$_2$Ti$_2$O$_7$, where uniaxial pressure along the $[110]$ direction was shown to lift the six-fold degeneracy of spin ice, splitting it into a lower-energy set of four and a higher-energy set of two configurations~\cite{Mito2007Mar}. This observation is consistent with the phenomenology captured by our model.

While in bulk spin ice one typically expects small perturbations, recent experiments on thin-film systems suggest that even sizeable modifications of the nearest-neighbor coupling might indeed be achievable. Growing spin-ice compound thin films on a substrate with a different lattice constant can hereby lead to epitaxial strain of the lattice. Depending on the growth direction of the spin ice compound on the growing substrate, the thin film can be strained in different directions~\cite{Bovo2014Mar, Bovo2019Mar, Barry2019Aug}. In epitaxially strained thin films of Dy$_2$Ti$_2$O$_7$, the strain effects have been estimated to shift $J_{\text{nn}}$ by $0.05$K - $0.5$K~\cite{Bovo2014Mar}, and the degeneracy of the six low-energy spin ice configurations has been estimated to split via the strain up to $0.55$ K~\cite{Bovo2019Mar}, corresponding to $4\delta$ in our model. This suggests the parameter range chosen in our simulations of Sec.~\ref{sec:Thermodynamics} ($\delta = 0.05 J$) is within an experimentally realistic range. 
 
While these results were obtained in uniform pyrochlore systems, our work also extended to the breathing pyrochlore lattice, where bond strengths can differ between the two tetrahedral sublattices. In a potential breathing pyrochlore material, it is conceivable that pressure applied along specific crystal directions could compress the two types of tetrahedra differently. The compression generally depends on the interaction strength between the sites within a crystal~\cite{Ashcroft1976Jan}. In the breathing pyrochlore, it is known that the interaction strength differs between sublattices, which motivates our assumption of different compression on the individual sublattices. 
In our investigated toy model, we assume the interaction between the spin pairs $\{ S_0^z, S_3^z\}$ and  $\{ S_1^z, S_2^z\}$ to differ when pressure is applied along $[001]$ or on the plane perpendicular to this direction.
Another possible route to achieve the staggered strain on the different tetrahedral sublattices could involve using epitaxially strained thin films grown on a breathing pyrochlore substrate. If the lattice mismatch between film and substrate differs between the two types of tetrahedra, this might, in principle, induce strain of opposite sign on the two sublattices. While this approach would require significant material engineering, it offers a potential path toward realizing the asymmetry in $\delta$ considered in our model Hamiltonian.

To summarize, realizing the anisotropic breathing spin ice model experimentally would require a (breathing) pyrochlore lattice in which the interactions on both sublattices match those of the Hamiltonian~\eqref{eq:Ham_Q}. Achieving the desired anisotropic bond dependence may be experimentally feasible through several routes, including sublattice-selective compression of single crystal breathiong pyrochlores via application of directional pressure, or epitaxial strain engineering using lattice-mismatched substrates.

\section{Omni-Plane projections and transition temperatures}
\label{sec:App_OmniPlane}

\begin{figure*}
    \centering
    \includegraphics[width = 1. \linewidth]{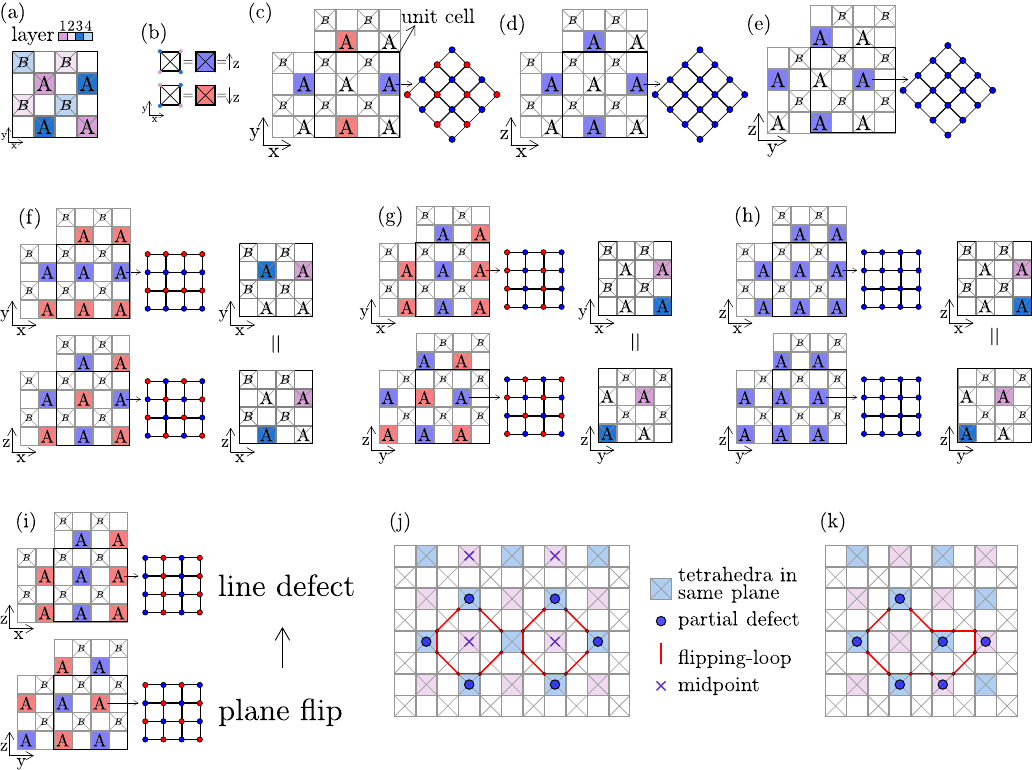}
    \caption{\textbf{Omni-Plane ground state projections onto different planes.} (a) The projection of a single conventional unit cell of the pyrochlore lattice onto the x,y-plane shows the four layers of tetrahedra as projected onto the plane. (b) The two ground-state configurations of an $A$ tetrahedron have a global magnetization vector pointing parallel or antiparallel to the global $z$ direction. This is shown by the projected tetrahedron-square having color blue or red. (c-e) Projection of the unit cell in a ground-state configuration onto different planes. Each layer of $A$ tetrahedra forms a square lattice, tilted by $45^\circ$ with respect to the unit cell. For the different ground-states in different planes, the resulting square lattice has $A$ tetrahedra being antiferromagnetically ($x,y$-plane) or ferromagnetically ($x,z$- and $y, z$-plane) aligned. (f-g) The three states that simultaneously fulfill two of the ground-state criteria in two perpendicular planes. The projection of the unit cell forms a square lattice with a stripe-pattern, an alternating-pattern, or a uniform pattern. On the right-hand side, the position of two $ A$-tetrahedra in different unit-cell projections is shown.  (i) Flipping a plane of $A$ tetrahedra in the $y,z$-plane generates another ground state, but creates a line defect in the uniform pattern of the $x,y$-plane. (j) Flipping flips in an octagon-loop shape within a single tetrahedra plane, creates partial defects at the sides of the octagon. Grouping together octagons creates partial defects at the outermost border. (k) In the \textit{omni-plane} phase the flipping-loops can also connect different planes.}
    \label{fig:Project_OP}
\end{figure*}

When $\delta_A<0$ and $\delta_B = 0$, then a ground state for the system is given by all configurations that have $\mathsf{Q}_A = 0$ and the partial charge $\mathsf{q}_A^{03} = -\mathsf{q}_A^{12} = \pm 2$, while on the $B$ sublattice only the total charge must be $\mathsf{Q}_B = 0$; the value of the partial charge is not fixed. We find that this constraints leads to three types of ground state constructions: \textit{(a)} in a plane perpendicular to the $z$ direction the $A$ tetrahedra are antiferromagnetically aligned, \textit{(b)} in a plane perpendicular to the $y$ direction the $A$ tetrahedra are ferromagnetically aligned, or, \textit{(c)} in a plane perpendicular to the $x$ direction the $A$ tetrahedra are ferromagnetically aligned. This structure is shown in Fig.~\ref{fig:Project_OP}(c-e) as projected onto the respective plane. 
Fulfilling requirement \textit{(a)} means that in $z$ directions the alignment of planes is independent, but within $x,y$-planes, the tetrahedra are correlated. The counting of the total number of states is straightforward: we start from a configuration \textit{(a)}, \textit{(b)}, or \textit{(c)} and count in how many different ways we can flip planes perpendicular to the specified direction. In direction $\alpha\in \{x, y, z\}$ there are $2 L_{\alpha}$ individual planes of $A$ tetrahedra. Therfore there are $\binom{2L_{\alpha}}{i}$ possible ways to flip $i$ planes of $A$ tetrahedra. Counting all possibilities would give us the total number of configurations $\Omega = \sum_{\alpha =x,y,z} \sum_{i=1}^{2L_{\alpha}} \binom{2L_{\alpha}}{i} = \sum_{\alpha =x,y,z}2^{2L_{\alpha}}$. However, here, six states are counted twice as explained below:

There are only two states that at the same time fulfill requirement \textit{(a)} and requirement \textit{(b)}: when in a plane perpendicular to the $z$ direction, the $A$ tetrahedra are antiferromagnetically aligned \emph{and} the magnetization of neighboring planes is alternating. One of these states is shown in Fig.~\ref{fig:Project_OP}(f) with the respective projection onto the plane. There are only two states that at the same time fulfill requirement \textit{(a)} and requirement \textit{(c)}: when in a plane perpendicular to the $x$ direction, the $A$ tetrahedra are ferromagnetically aligned \emph{and} the magnetization of neighboring planes is alternating. One of these states is shown in Fig.~\ref{fig:Project_OP}(g) with the respective projection onto the plane. There are only two states that at the same time fulfill requirement \textit{(b)} and requirement \textit{(c)}: when in a plane perpendicular to the $y$ direction, the $A$ tetrahedra are ferromagnetically aligned \emph{and} all the planes' magnetizations are equally aligned. One of these states is shown in Fig.~\ref{fig:Project_OP}(h) with the respective projection onto the plane. This, in total, gives six states that are counted twice when considering all possible configurations that can be generated by the plane flips.
To account for this double counting, we subtract those states once from our calculations to reach $\Omega = \sum_{\alpha =x,y,z} \sum_{i=1}^{2L_{\alpha}-1} \binom{2L_{\alpha}}{i} = \sum_{\alpha =x,y,z}(2^{2L_{\alpha}}-2)$. 

To understand the spin structure factor (SSF) shown in Fig.~\ref{fig:SSF_SF}, we can imagine starting from any of the six overlapping ground states. The projections onto the three planes ($x,y$, $x,z$, and $y,z$) of the overlapping ground states are shown in Fig.~\ref{fig:Project_OP}(f-h). In the $x,y$-plane, we either have a horizontally or a vertically striped pattern on a square lattice. This would lead to Bragg peaks at $(0,\pm1)$ or $(\pm1, 0)$ in the square lattice. In the $x,z$-plane and the $y,z$-plane, we either have ferromagnetic or antiferromagnetic alignment on a square lattice Brillouin zone. This would lead to a Bragg peak at $(0,0)$ or $(\pm1, \pm1)$ in the square lattice Brillouin zone. Flipping any plane of a tetrahedron starting from any of these states will introduce a line of defects into the square lattice projection in the plane perpendicular to the flipped plane. This is shown with the example of Fig.\ref{fig:Project_OP}(i). The line defects lead to a broadening of the Bragg peaks along the line-defect direction. This leads to an elongated peak at $(\pm1, 0, 0)$ and $(0, \pm1, 0)$ from the stripe order in the $[hl0]$ (or $x,y$-) plane. On the other hand, the Bragg peak from the ferromagnetic order at $(0,0,0)$ is elongated horizontally, and the Bragg peaks from the antiferromagnetic order at $(\pm1,0,\pm1)$ are elongated vertically. This created the appearance of a square grid. 
\\

To estimate the critical temperature of the omni-plane and plane-paramagnetic phase, we employed a Peierls-type argument~\cite{Peierls1936Oct} based on the energy cost and entropy of defect loops.

We can first analyze the argument in the \textit{plane-paramagnetic} phase. When the planes decouple, the $B$ tetrahedra form an antiferromagnetic square lattice. By forming octagon-shaped loops between four neighboring $B$ tetrahedra in a single plane, we can create four partial changes in the $B$ tetrahedra touched by the octagon, each with an energy difference of $\Delta E = 4 \delta$. An octagon loop connecting four tetrahedra in a common plane is shown in Fig.\ref{fig:Project_OP}(j). This is the equivalent of thinking about flipping a single spin (in the middle of the octagon) in the square lattice and creating four frustrated bonds (the four defects on the $B$ tetrahedra). Linking together octagons creates $L$ defects at the outermost boundary of the \textit{updated} area with perimeter $L$. The possibilities to construct an updated area with perimeter $L$ may be named $\mu$. From this, we get 
$$\Delta F = \Delta E - T\Delta S = L \cdot 4\delta - T \cdot L \ln{\mu}. $$
An upper bound for $\mu$ can be found to be $\mu < 3$~\cite{Lieb2005May}, easily seen if considering the possibilities of continuing a path on a square lattice without backtracking. 
This leads to a critical temperature of $T_c = \frac{4\delta}{\ln{\mu}}$. From Onsager's solution of the square lattice Ising model~\cite{Onsager1944Feb} it is known, that $\mu_{\text{square}} = 1+\sqrt{2}$. Since there are, however, two possibilities to create this \textit{dual} square lattice by the midpoints of the octagon loops, we expect the exact value for the entropy to be $S = 2\cdot \mu^L$ and therefore the critical temperature $T_c = \frac{4\delta}{\ln{(1+\sqrt2)} +\frac{\ln{2}}{L}}$, tending towards the Onsager solution for larger system sizes.\\

In the \textit{omni-plane} phase, we can apply a similar principle to the aforementioned derivation. However, here we have the option to extend the loop also out of a single correlated plane. Connecting two planes with a loop is shown in Fig.\ref{fig:Project_OP}(k). The $A$ tetrahedra are (anti)ferromagnetically aligned in planes perpendicular to one of the three spatial directions. The octagon loops within these correlated planes, again, generate four defects at the corners (each with energy difference $\Delta E = 4 \delta_B$). We can, however, also extend this loop into planes above or below by flipping an appropriate oppositely aligned spin on a crossing $B$ tetrahedron. This increases the number of possible paths available for a loop and gives an approximate upper bound for $\mu \le 6$. This leads to the critical temperature $T_c \ge \frac{4\delta}{\ln{6}}$.
For the ferromagnetic case, the explanation is given in the text.

\section{Local spin basis}
\label{Sec:App_LocalSpinBasis}
Our interaction Hamiltonian \eqref{eq:Ham_Q} resides on the breathing pyrochlore lattice. The unit cell of the pyrochlore lattice corresponds to a face-centered cubic (fcc) lattice with an $A$ tetrahedron at each lattice point. The four fcc lattice sites are denoted by $\mathbf{R}_i$ and are located at positions
\begin{equation}
    \begin{split}
        \mathbf{R}_0 &= (100), \qquad
        \mathbf{R}_1 = (1\frac{1}{2}\frac{1}{2}),\\
        \mathbf{R}_2 &= (\frac{1}{2}\frac{1}{2}0), \qquad
        \mathbf{R}_3 = (\frac{1}{2}0\frac{1}{2}).\\
    \end{split}
\end{equation}

At each fcc lattice site, an $A$ tetrahedron resides, where the four spins are located at positions $\mathbf{r}_i$:

\begin{equation}
    \begin{split}
        \mathbf{r}_0 &= \frac{1}{4} (100), \qquad
        \mathbf{r}_1 = \frac{1}{4} (001), \\
        \mathbf{r}_2 &= \frac{1}{4} (111), \qquad
        \mathbf{r}_3 = \frac{1}{4} (010). \\
    \end{split}
    \label{coordinates}
\end{equation}

The position of a spin $S_{i,\alpha}^z$ in the lattice is given by $\mathbf{R}_{i,\alpha} = \mathbf{R}_i + \mathbf{r}_{\alpha}$, where $i$ is the site index, and $\alpha$ is the spin index. The distance between two spins is then defined as $\mathbf{R}_{i \alpha, j \beta} = \mathbf{R}_{i \alpha} - \mathbf{R}_{ j \beta}$.
We use the local spin basis, $\hat{S}_{i,\alpha}^z = S_{i,\alpha}^z \cdot \mathbf{e}_{\alpha}$, where each spin residing at position $\alpha$ in a tetrahedron located at site $i$ of the fcc unit cell, is either aligned or anti-aligned with $\mathbf{e}_{\alpha}$. The unit vector $\mathbf{e}_{\alpha}$ vectors pointing from the center of a single tetrahedron into one of the four corner points:
\begin{equation}
\label{eq:local_basis}
    \begin{split}
        \mathbf{e}_0 &= (-1,1,1)/\sqrt{3}\\
        \mathbf{e}_1 &= (1,1,-1)/\sqrt{3}\\
        \mathbf{e}_2 &= (-1,-1,-1)/\sqrt{3}\\
        \mathbf{e}_3 &= (1,-1,1)/\sqrt{3}.
    \end{split}
\end{equation}

The Hamiltonian can now be written as 
\begin{equation}
\begin{split}
\label{eq:H_with_R}
        H &= \sum_{\braket{(i \alpha, j \beta)}} J^{i \alpha, j \beta} S_{i,\alpha}^z S_{j,\beta}^z\\
        &= \sum_{\braket{\alpha, \beta}, i=j} J^{i \alpha, i \beta} S_{i,\alpha}^z S_{i,\beta}^z + \sum_{\braket{\alpha, \beta}, i\neq j} J^{i \alpha, j \beta} S_{i,\alpha}^z S_{j,\beta}^z,
\end{split} 
\end{equation}
where we distinguish between bonds within $A$ tetrahedra ($\sum_{\braket{\alpha, \beta}, i=j}$) and bonds that constitute $B$ tetrahedra ($\sum_{\braket{\alpha, \beta}, i\neq j}$).
Here, the interaction matrix is
\begin{equation}
\label{def:J_ij_ab}
    \begin{split}
        J^{i\alpha, i \beta} &= J_A \begin{pmatrix}
            0&1&1&1\\ 1&0&1&1\\1&1&0&1\\1&1&1&0
        \end{pmatrix} + 
        \delta_A \begin{pmatrix}
            0&0&0&1\\ 0&0&1&0\\0&1&0&0\\1&0&0&0\end{pmatrix} + \bar{c} \mathbb{I} \\
        J^{i\alpha, j \beta} &= J_B \begin{pmatrix}
            0&1&1&1\\ 1&0&1&1\\1&1&0&1\\1&1&1&0
        \end{pmatrix}\delta(|\mathbf{R}_{i \alpha, j \beta}|- r_{nn}) \\
        &+ 
        \delta_B \begin{pmatrix}
            0&0&0&1\\ 0&0&1&0\\0&1&0&0\\1&0&0&0\end{pmatrix}\delta(|\mathbf{R}_{i \alpha, j \beta}|- r_{nn})+  \bar{c} \mathbb{I},
    \end{split}
\end{equation}
where $r_{nn} = \frac{\sqrt{2}}{4}$ is the nearest neighbor distance.

\section{Self-consistent Gaussian approximation}
\label{sec:App_MFT}
The self-consistent Gaussian approximation is based on the softening of the hard spin constraint $|\mathbf{S}_i| = S^2$ to a soft one with $\braket{\mathbf{S}_i} = S^2$~\cite{Reimers1991Jan, Garanin1999Jan, Canals2002Apr, Isakov2004Oct, McClarty2014May, Conlon2010Jun}. This constraint is enforced energetically with a Lagrange multiplier $\lambda$, such that 
\begin{equation}
\label{eq:beta_H}
    \beta \tilde{H} = \sum_{\langle i,j\rangle \in A,B} (J_{A,B}^{ij} + \lambda \delta_{ij}) S_i^z S_j^z.
\end{equation}
The spin length is now energetically fixed self-consistently through
\begin{equation}
    \braket{S_{\alpha}^2} = \frac{1}{4 L^3} \sum_{\mathbf{q} \in \text{BZ}} \text{Tr}\left[ \lambda \mathbb{I} + \beta \hat{J}(\mathbf{q}) \right]^{-1}.
\end{equation}

To calculate this and the structure factor in the SCGA, we need to perform the Fourier transform of the Hamiltonian \eqref{eq:H_with_R}.
This gives 
\begin{equation}
    H' = \sum_{\mathbf{q} \in \text{BZ}} \sum_{\alpha, \beta} J^{\alpha \beta}(\mathbf{q}) S_{\alpha}^z(\mathbf{q}) S_{\beta}^z(-\mathbf{q}),
\end{equation}
where $J^{\alpha \beta}(\mathbf{q})$ is the Fourier transform of the interaction matrix \eqref{def:J_ij_ab}.
This is now defined as
\begin{equation}
    J^{\alpha \beta}(\mathbf{q}) = \frac{1}{N} \sum_{\mathbf{R}_{i \alpha, j \beta}} J^{i \alpha, j \beta} e^{-i \mathbf{q} \cdot \mathbf{R}_{i \alpha, j \beta} }.
\end{equation}
Inserting the definition of $J^{i\alpha, i \beta}$ \eqref{def:J_ij_ab} and the coordinates for $\mathbf{R}_{i \alpha, j \beta}$ for $i = j$ and $i \neq j$ gives the interaction matrix in \eqref{def:J_ab_q}.

\begin{equation}
\label{def:J_ab_q}
    J^{\alpha \beta} (\mathbf{q}) = \tilde{c} \cdot \mathbb{I}_4 + \cdot \begin{pmatrix}
        0 &c_{01} & c_{02} &c_{03} \\
        \overline{c_{01}} & 0 & c_{12} &c_{13} \\
        \overline{c_{13}} &\overline{c_{12}} & 0 &c_{23} \\
        \overline{c_{03}} &\overline{c_{13}} & \overline{c_{23}}&0 \\
    \end{pmatrix},
\end{equation}
where the coefficients are
\begin{equation}
    \begin{split}
        c_{01} & = J_A e^{-i (q_x-q_z)} + J_B e^{+i (q_x -q_z)} \\
        c_{02} & = J_A e^{-i (-q_y-q_z)} + J_B e^{+i (-q_y-q_z))} \\
        c_{03} & = (J_A+ \delta_A) e^{-i (q_x-q_y)} + (J_B+ \delta_B) e^{+i (q_x -q_y)} \\
        c_{12} & = (J_A+ \delta_A) e^{-i (-q_x-q_y)} + (J_B+ \delta_B) e^{+i (-q_x -q_y)} \\
        c_{13} & = J_A e^{-i (-q_y+q_z)} + J_B e^{+i (-q_y+q_z)} \\
        c_{23} & = J_A e^{-i (q_x+q_z)} + J_B e^{+i (q_x+q_z)},
    \end{split}
\end{equation}
and $\overline{c_{\alpha \beta}}$ is the complex conjugate of $c_{\alpha \beta}$.

Note that $\tilde{c}$ in Eq.~\eqref{def:J_ab_q} marks a constant energy shift that can be chosen freely. Here, we choose $\tilde{c}$ such that the minimum eigenvalue of $J(\mathbf{q})$ is zero. This also makes sure that the Lagrange multiplier $\lambda$ only takes values between one and zero, making a physical interpretation possible.  

Since with a basis transformation, Eq.~\eqref{eq:beta_H} can be written as
\begin{equation}
    \beta \tilde{H} = \sum_{\mathbf{q}, n} (\beta \varepsilon_n(\mathbf{q})+ \lambda) |\sigma_n(\mathbf{q})|^2,
\end{equation}
where $\varepsilon_n(\mathbf{q})$ are the eigenvalues of $J(\mathbf{q})$ and $\sigma_n(\mathbf{q})$ are the ``normal modes" of the system~\cite{Chung2022Mar}.
With this, $\lambda^{-1}$ can be interpreted as the \textit{thermal occupation value} for the lowest energy mode, since
\begin{equation}
    \braket{|\sigma_n(\mathbf{q})|^2} = \frac{1}{\lambda + \beta \varepsilon_n(\mathbf{q})}.
\end{equation}
With our convention of $\tilde{c}$, we have set the energy of the lowest energy mode exactly equal to zero, such that the thermal occupation value of this mode is exactly $\lambda^{-1}$.

We can now discuss some limiting cases for $\lambda$:
At high temperatures, $\beta \rightarrow0$, all energy modes are equally singly occupied and $\lambda= 1$.
For the \textit{isotropic} nearest neighbor spin ice, as $\beta \rightarrow \infty$, the lowest two energy modes become flat and fully degenerate and $\lambda \rightarrow\frac{1}{2}$. This tells us that the high-energy modes depopulate, and the degenerate low-energy modes are doubly occupied. For the \textit{planar ice}, as $\beta \rightarrow \infty$, only one lowest energy mode becomes flat and $\lambda \rightarrow\frac{1}{4}$.
If a normal mode becomes critical at any temperature, then $\lambda^{-1} \rightarrow\infty$ or $\lambda \rightarrow0$. This indicates a phase transition with long-range order in the SCGA. 
\begin{figure}[h!]
    \centering
    \includegraphics[width = 1 \columnwidth]{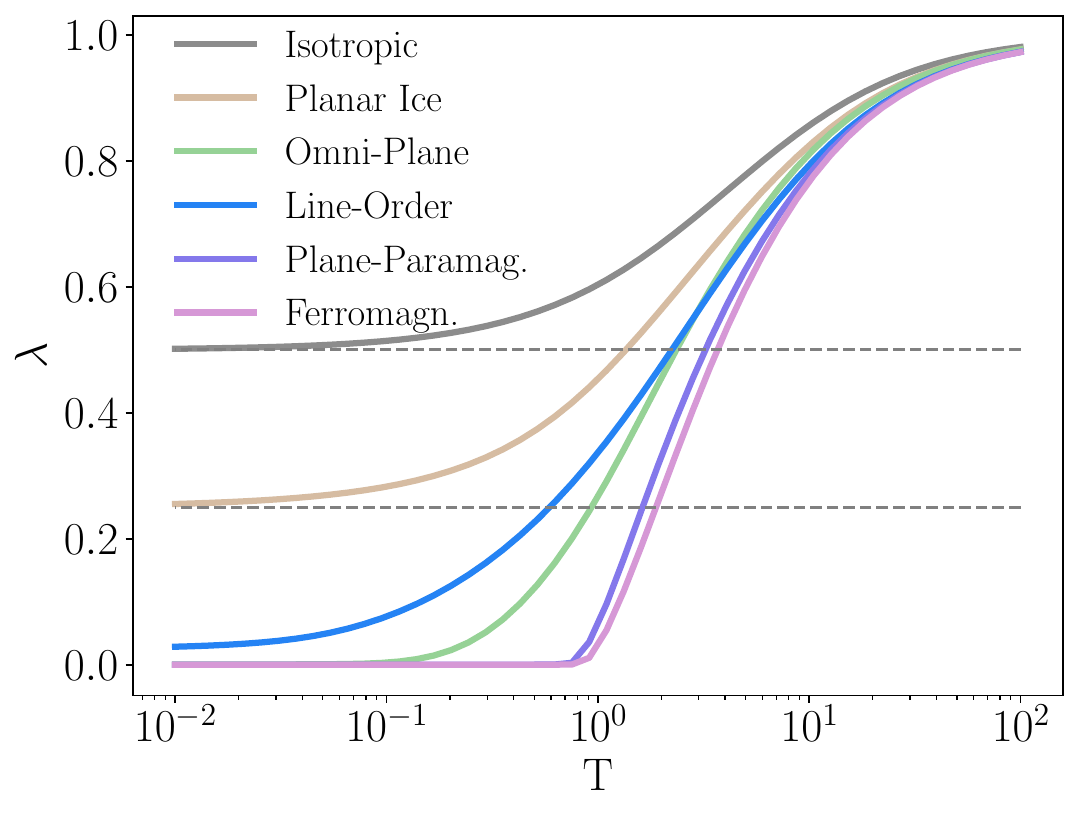}
    \caption{\textbf{Lagrange-multiplier for the SCGA.} Lagrange-multiplier that fixed the spin constraint in the self-consistent Gaussian approximation. For \textit{isotropic} spin ice, $\lambda\rightarrow\frac{1}{2}$ as $T\rightarrow0$. For \textit{planar ice}, $\lambda \rightarrow\frac{1}{4}$ as $T \rightarrow0$. For the \textit{line-order} phase, $\lambda\rightarrow0$ as $T \rightarrow 0$, suggesting long-range order at $T = 0$. For the three remaining phases, the sudden drop of $\lambda \rightarrow0$ at a finite temperature indicates a finite-temperature phase transition.   }
    \label{fig:lambda_all}
\end{figure}

The Lagrange multiplier for the six phases is shown in Fig.~\ref{fig:lambda_all}. We can observe the expected behaviour as $T\rightarrow0$ for \textit{isotropic} spin ice ($\lambda \rightarrow \frac{1}{2}$) and for \textit{planar ice} ($\lambda \rightarrow \frac{1}{4}$). For the \textit{line-order} phase, we can observe how $\lambda$ monotonically decreases as $T$ approaches 0. This suggests a transition to long-range order to appear only at $T=0$, similarly to what is expected in a one-dimensional Ising chain~\cite{Ising1925Feb}. By contrast, the Lagrange multiplier of \textit{omni-plane} phase, the \textit{plane-paramagnetic} phase, and the \textit{ferromagnetic} phase seem to have a sudden drop to zero at a finite temperature. It is important to note, however, that $\lambda$ can only really reach zero at a finite temperature for $L\rightarrow\infty$. While for the \textit{plane-paramegnetic} phase and the \textit{ferromagnetic} phase, the drop in $\lambda$ appears sudden and steep, $\lambda$ of the \textit{omni-planar} phase approaches small values more gradually. A finite-size scaling analysis of $\lambda$ in the \textit{omni-planar} phase, however, still points towards a finite-temperature phase transition within the SCGA. This supports the numerical findings in Sec.~\ref{Sec:Cv_anisotropic}, indicating a finite-temperature phase transition in the three phases with any $\delta_i<0$.

\begin{figure}
    \centering
    \includegraphics[width=1\linewidth]{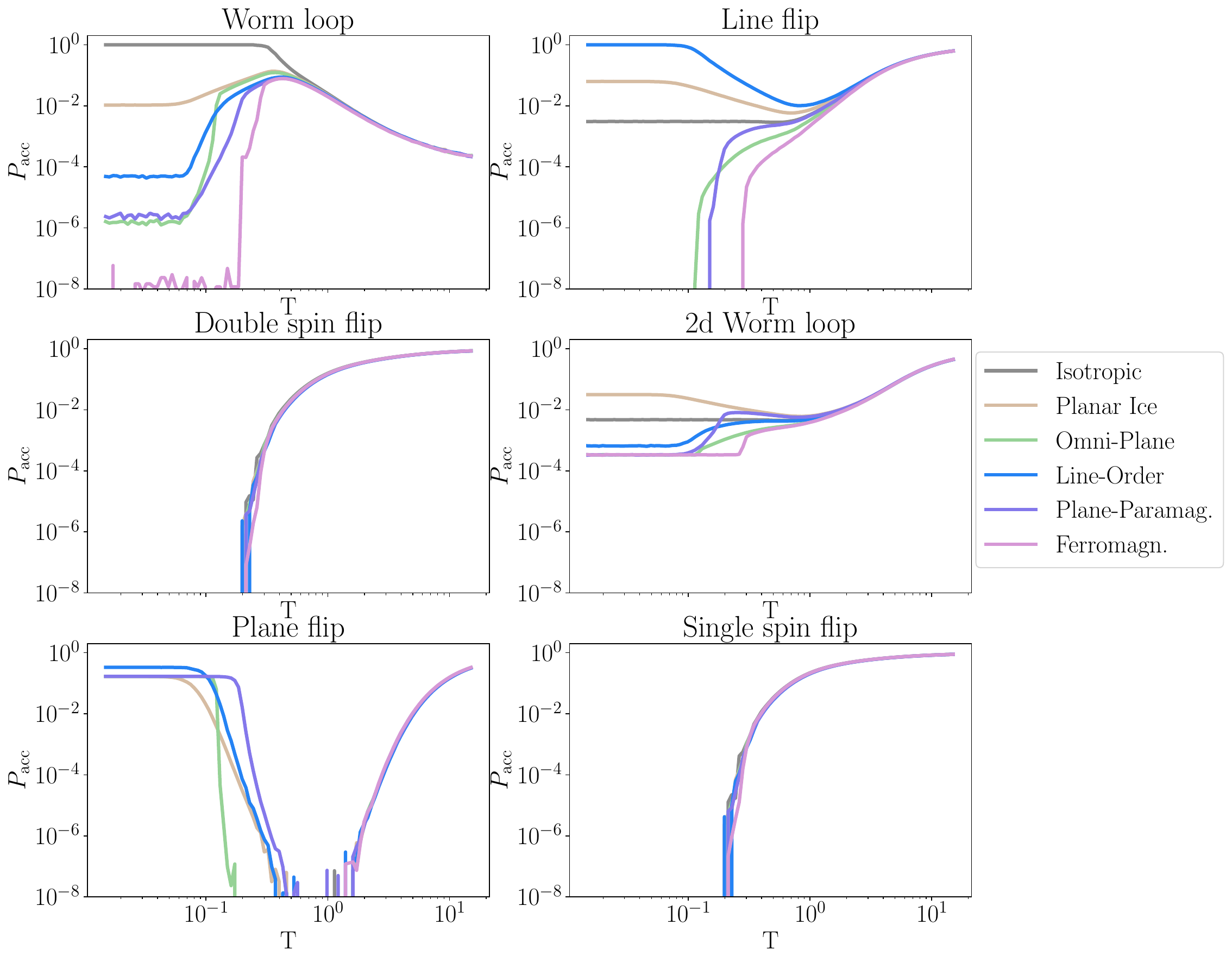}
    \caption{\textbf{Acceptance rates.} Comparison of the acceptance rate ($P_{\mathrm{acc}}$) for different update schemes.}
    \label{fig:App_acc_rates}
\end{figure}

\section{Monte-Carlo simulation}
\label{sec:App_MonteCarlo}

All numerical data presented in this work was obtained using Monte-Carlo simulations. Specifically, the data shown in Fig.~\ref{fig:Cv_S} was generated for a system of size $L_x \times L_y \times L_z = 64$ face-centered cubic (fcc) unit cells, with each unit cell containing $16$ spins, resulting in a total of $N_{\text{spin}} = 1024$ spins. Periodic boundary conditions were applied in all three spatial directions ($x$, $y$, and $z$) to minimize boundary effects.  

During each Monte-Carlo time step, six types of updates were employed to enhance equilibration: single spin flips, worm loop updates, and other cluster-updates (see App.~\ref{Sec:App_ClusterUpdates}). 
For each measurement temperature, the system was evolved until $N_{\text{sweeps}} = N_{\text{spin}} \cdot 10^3$ accepted updates (including spin flips and worm loops) were performed. Measurements were taken at equilibrium.

To generate data across different temperatures, the system was progressively cooled in 10 steps from the last measurement temperature. At each temperature step, the system was equilibrated before proceeding to the next step. Finally, at the next measurement-temperature, an additional equilibration phase was conducted. 

Here again, the next sweep measurement was performed.

\section{Cluster update schemes}
\label{Sec:App_ClusterUpdates}
For isotropic nearest-neighbor spin ice, in numerical simulations, typically updates are given by single spin flips as well as worm loop updates~\cite{Barkema1998Jan, Isakov2004Sep, Sandvik2006Apr}. Since at low temperatures, spin freezing makes equilibration times with only single-spin flips extremely long, the worm loop algorithm finds update moves at a zero energy cost~\cite{Melko2001Jul}.
The ground state degeneracy of all anisotropic systems is, however, greatly reduced compared to the isotropic spin ice case. This also leads to an increase in equilibration time in numerical simulations when only employing the two aforementioned update schemes. 
The problem lies in the acceptance rate of the attempted updates. A measurement series at a particular temperature is only completed if a prefixed number of spin updates are accepted. 
For isotropic spin ice below the crossover, where the system is in a ground state configuration, all attempted worm loop updates are accepted. However, this acceptance rate drops significantly for all anisotropic systems. Now, to achieve the same number of accepted updates, either the number of attempted updates must be increased significantly, or a new type of update with a higher acceptance rate must be proposed. The latter is oped for to produce the data in Fig.~\ref{fig:Cv_S} and the particular update schemes are discussed here.

\textit{Single spin flip}: A single spin is randomly selected and flipped with the Metropolis probability $\min[1, \exp(-\Delta E / T)]$, where $\Delta E$ is the change in energy associated with the spin-flip.  

The additional update schemes rely on the operator structure to generate all ground states of the anisotropic phases. 
As discussed in the main text, the \textit{worm loop} relies on the fact that the ground state constraint in isotropic spin ice consists of two spins pointing into a tetrahedron and two spins pointing out. Finding any closed loop of spins pointing ``\textit{in}" and ``\textit{out}" alternately will always conserve the ground state constraint, and a flip will lead to another ground state.
\textit{Worm loop update}: A closed loop of spins aligned head-to-tail is identified, starting from a randomly chosen spin. The spins within the loop are flipped collectively with the same Metropolis probability. The loop construction ensures that no monopoles are created, annihilated, or moved during the update, although partial charges may be created or annihilated.  

\begin{figure}
    \centering
    \includegraphics[width=1\columnwidth]{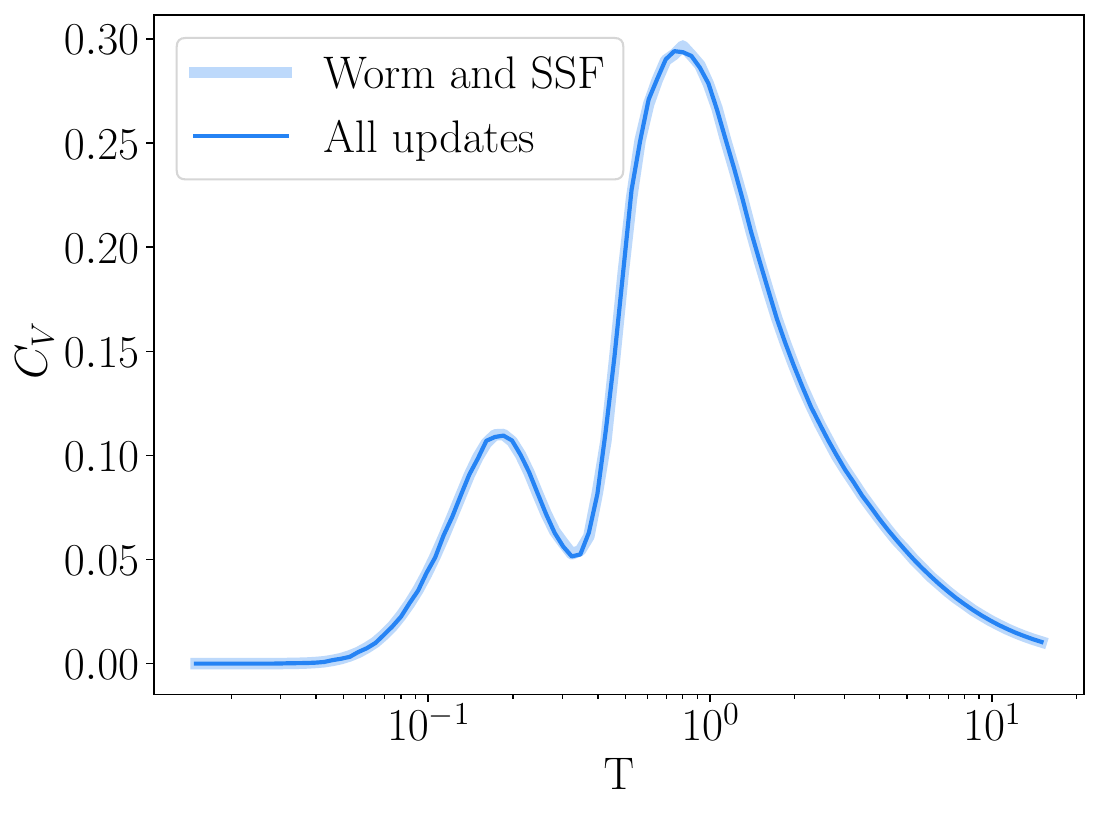}
    \caption{\textbf{Comparison between update schemes.} Specific heat data for line-order phase with $\delta_A = \delta_B = 0.05$ and $L=2$ obtained by only using worm loop updates and single spin flips (SSF) or all update schemes. The curves perfectly overlap in the whole temperature regime. }
    \label{fig:App_update_schemes}
\end{figure}
A \textit{Line flip} consists of randomly choosing a line of spin pairs (either $\{S_0^z, S_3^z\}$ or $\{S_1^z, S_2^z\}$, see Fig.~\ref{fig:Lines_Planes}) and flipping it. Detailed balance is fulfilled, as the probability of choosing any line is $W_{\mathrm{choice}}(i\rightarrow j) = \frac{1}{N_{\mathrm{lines}}}$, where $N_{\mathrm{lines}}$ is the number of such spin lines in the lattice and therefore independent of the current configuration. The acceptance probability of flipping the line is given by the Metropolis criterion $A(i\rightarrow j) = \mathrm{min}[1, e^{-\beta (E_j-E_i)} ]$. \\
A \textit{Double spin flip} consists of choosing randomly a spin and any adjacent spin and attempting to flip them both. The acceptance probability is given by the Metropolis probability.\\
A \textit{2d worm loop} consists of forming a closed loop in a single $x,y$-plane of $B$ tetrahedra. This means on $A$ tetrahedra, spins can only be flipped as spin-pairs $\{S_0^z, S_3^z\}$ or $\{S_1^z, S_2^z\}$, while on the $B$ tetrahedra the choice is free.\\
A \textit{Plane flip} consists of flipping a plane of either $A$ or $B$ tetrahedra with a common coordinate. This can be a plane perpendicular to any special direction. \\
In Fig.~\ref{fig:App_acc_rates}, the acceptance rates for the different update schemes are shown. It is evident that the acceptance rate for single or double spin flip drops much below $10^{-8}$ for temperatures below the first crossover. 
The four new update schemes help to increase the mean acceptance rate for all phases above $10^{-2}$.

The underlying ordering processes are not changed by the introduction of the new cluster-update schemes, but rather, achieving the result numerically is merely accelerated. This is shown exemplarily in Fig.~\ref{fig:App_update_schemes} for a system size of $L_x \times L_y \times L_z =8$, where the data obtained by only using the worm loop updates and single spin flips, and all six cluster update schemes detailed above are shown. The two curves overlap fully.

\bibliography{Bib}

\end{document}